\documentclass[12pt]{article}

\usepackage{amsmath}
\usepackage{amssymb}
\usepackage{latexsym}
\usepackage{graphicx}
\usepackage{psfrag,fancyhdr,epsfig}

\begin{document}

\begin{titlepage}

\vspace*{1cm}
\begin{center}
{\bf \Large Brane Decay of a (4+n)-Dimensional Rotating\\[1mm] Black Hole: spin-0 particles}

\bigskip \bigskip \medskip

{\bf G. Duffy}$^1$, {\bf C. Harris}$^2$, {\bf P. Kanti}$^3$ and
{\bf E. Winstanley}$^4$

\bigskip
$^1$ {\it Department of Mathematical Physics, University
College Dublin,\\ Belfield, Dublin 4, Ireland}

$^2$ {\it Cavendish Laboratory, University of Cambridge, \\Madingley Road,
Cambridge CB3 0HE, United Kingdom}

$^{3}$ {\it Department of Mathematical Sciences, University of Durham,\\
Science Site, South Road, Durham DH1 3LE, United Kingdom}

$^4$ {\it Department of Applied Mathematics, The University of Sheffield,\\
Hicks Building, Hounsfield Road, Sheffield S3 7RH, United Kingdom}

\bigskip \medskip
{\bf Abstract}
\end{center}
In this work, we study the `scalar channel' of the emission of Hawking radiation
from a $(4+n)$-dimensional, rotating black hole on the brane. We numerically
solve both the radial and angular part of the equation of motion for the scalar
field, and determine the exact values of the absorption probability and of the
spheroidal harmonics, respectively. With these, we calculate the particle, energy
and angular momentum emission rates, as well as the
angular variation in the flux and power spectra -- a distinctive feature of emission
during the {\it spin-down phase} of the life of the produced black hole. Our analysis
is free from any approximations, with our results being valid for arbitrarily large
values of the energy of the emitted particle, angular momentum of the black hole and dimensionality of spacetime. We finally compute the total emissivities for the
number of particles, energy and angular momentum and compare their relative
behaviour for different values of the parameters of the theory.

\end{titlepage}

\section{Introduction}

The formulation of the theory with Large Extra Dimensions \cite{ADD} was driven by the
motivation to address the hierarchy problem (for some early works, see \cite{early}).
While all Standard Model fields are restricted to live on a (3+1)-dimensional brane,
gravity is allowed to propagate in the $(4+n)$-dimensional bulk; above the scale of
decompactification of extra dimensions $M_*$, that can be as low as 1 TeV, gravitational
interactions become strong and thus comparable to the Electroweak interactions.
Moreover, above $M_*$, all effective theories cease to be valid, and a more fundamental
theory, that describes all forces including gravity, must take over. In the context
of such a theory, transplanckian collisions of particles become particularly
important as their product cannot be ordinary particles any more but rather
heavy objects. This opens up the exciting possibility of observing strong
gravitational phenomena, linked to a Quantum Theory of Gravity, at a low energy
scale, possibly at the TeV scale.

The creation of mini black holes as the result of such transplanckian particle
collisions \cite{creation} has drawn considerable interest during the last few years.
Such black holes may be created either at colliders \cite{colliders} or in high
energy cosmic-ray interactions \cite{cosmic} (for an extensive discussion of the
phenomenological implications and a more complete list of references, see the reviews
\cite{Kanti, reviews, Harris}). If the produced black holes have a mass considerably
larger than the fundamental Planck mass $M_*$, quantum gravity effects can be safely
ignored, and the black holes can be treated as classical objects. Their most important
characteristic, as well as the most prominent signature of their creation, will
be the emission of Hawking radiation in the form of elementary particles. This will
take place both in the bulk and on the brane, with the latter `channel' being the most
important from the phenomenological point of view. After shedding all additional
quantum numbers during a short {\it balding phase}, as dictated by the no-hair theorem
of General Relativity, the produced black hole will settle down to a Kerr-like
phase, the {\it spin-down phase}, during which the black hole will mainly lose its
angular momentum through the emission of Hawking radiation and superradiance.
After that, the {\it Schwarzschild phase} will commence with the
black hole (now spherically symmetric)
emitting Hawking radiation and gradually losing its actual mass.

The Schwarzschild phase of the life of a small higher-dimensional black hole formed
in a flat background has been admittedly the most well studied in the literature.
The task of determining the spectrum of Hawking radiation during this phase has been
attacked both analytically \cite{kmr1, Frolov1, kmr2} and numerically \cite{HK1}.
While the first set of works led to the derivation of useful, analytical formulae
for the emission rate, the latter work provided exact, numerical results valid
at all energy regimes. Both approaches reached the conclusion that the Hawking
radiation spectrum strongly depends on the existence of additional spacelike
dimensions in nature, with this dependence being reflected both in the number of
particles and energy emitted by the black hole per unit time as well as in the actual
type of particles emitted (scalars {\it vs.} fermions {\it vs.} gauge bosons).
Further studies have
also shed light on the dependence of Hawking radiation from a higher-dimensional
black hole on higher-derivative gravitational (Gauss-Bonnet) terms \cite{Barrau},
the mass of the emitted particles \cite{Jung-mass, Doran}, the charge of the black
hole \cite{Jung-charge}, and last but not least, the cosmological constant \cite{BGK},
with the latter leading to a distinct signature of its existence at the low-energy
part of the radiation spectrum.

The progress in studying the emission of Hawking radiation during the spin-down
phase of a small, higher-dimensional black hole has been much slower. This was
due to the more complicated gravitational background but also to additional technical
difficulties in solving the equation of motion for a particle propagating in such a
background. Analytical formulae for the emission rate were derived \cite{Frolov2,
IOP1} but the results were partial, being valid only at the low-energy regime,
for low angular momentum of the black hole, and for a specific dimensionality of
spacetime. The first exact numerical results for the Hawking radiation emission rate
in the form of scalar fields, during the spin-down phase, as well as for the
amplification due to superradiance, were presented in \cite{HK2}. These were valid
for arbitrary values of the energy of the emitted particle and for arbitrary values
of the angular momentum parameter of the black hole. That was the result of the
determination of the exact value of the angular eigenvalue -- being also involved
in the radial part of the equation of motion of the particle -- via numerical means
instead of the use of an approximate analytical formula. The analysis in \cite{HK2}
demonstrated that the suppression of the energy emission rate at the low-energy
regime as the angular momentum of the black hole increases, as was found in
\cite{IOP1}, is in fact overturned in the intermediate- and higher-energy regime
where a strong enhancement appears instead. In the case of superradiance, these
results confirmed the existence of the effect in the higher-dimensional case,
and demonstrated the exact energy amplification: although strongly mode-dependent,
this amplification can reach approximately a 10\% magnitude\,\footnote{Results on
the superradiance effect in the case of a higher-dimensional black hole were
also presented in \cite{IOP-proc} and \cite{Jung-super}.}. Shortly afterwards,
another study appeared in the literature \cite{IOP2}, focused on the study of the power
(energy) spectrum, as well as its angular distribution in the 5-dimensional case, but
once more the analysis relied on the assumption of low energy and low angular momentum.
Two more studies focused on 5-dimensional rotating black holes appeared recently
\cite{Nomura, Jung-rot}, where the question of the emission in the bulk was also
addressed.

At present, a comprehensive study of the emission of Hawking radiation from a
higher-dimensional black hole in its spin-down phase, that could provide exact
numerical results for arbitrary values of the angular momentum of the black hole,
 the energy of the emitted particle and the dimensionality of spacetime,
is still missing from the literature. Apart from the power (energy) spectrum,
whose study needs to be completed for arbitrary values of the aforementioned
parameters, the question of the emission rate of particles (flux spectrum) and
of the rate of loss of angular momentum of the rotating black hole also needs
to be addressed.
Moreover, the angular distribution of the emitted radiation -- a distinctive
signature of emission from a rotating higher-dimensional black hole -- needs
to be investigated, with the results again being free from any restrictive
assumption about the values of the fundamental parameters of the theory. It is
these tasks that we have undertaken to fulfil in this work. The strongly spin-dependent
techniques needed to be applied for the calculation of the spectra for different
types of particles, and the plethora of results that need to be derived in each
case, have forced us to restrict our attention in this work to the case of scalar
fields emitted by the black hole, and present the corresponding analysis and results
for higher-spin fields in a subsequent work \cite{CKW}.

Before presenting the outline of our paper, a brief discussion of the assumptions made
during our analysis should be added here. As mentioned earlier, it will be assumed
that the black hole mass $M_{BH}$ is considerably larger than the fundamental scale
of gravity $M_*$ in order to ensure that quantum effects are small and that the black hole geometry
can be safely considered as a classical object. The same constraint guarantees the
absence of any significant back reaction to the gravitational background due to the
change in the black hole mass after the emission of a particle: the assumption that
the temperature of the black hole, around which the emission spectrum is centered,
is much smaller than the black hole mass translates again to $M_{BH} \gg M_*$. We are
also going to assume that the horizon of the black hole $r_h$ is much smaller than
the (common) size $L$ of the additional compact spacelike dimensions so that the
black hole can be considered as embedded in a $(4+n)$-dimensional, non compact, empty
spacetime. Finally, since the value of the brane self-energy can be naturally assumed
to be of the order of the fundamental Planck scale $M_*$, and thus much smaller than
the black hole mass, its effect on the gravitational background can be ignored.

We will start, in Section 2, by presenting the theoretical framework for our analysis
including basic formulae for the gravitational background, the equation of motion for
the propagation of a scalar field in it, and the Hawking radiation emission spectra.
In section 3, a description of the numerical methods and techniques used for the determination
of the exact value of the angular eigenvalue and of the angular and radial part
of the wavefunction of the field will be given. After the angular eigenvalue is determined,
the radial part of the equation of motion is numerically solved to determine the
absorption probability (greybody factor) and thus the various spectra. In section 4,
we present our results for the flux, power and angular momentum spectra (integrated
over the azimuthal angle $\theta$) for arbitrary values of the energy of the particle,
the angular momentum of the black hole, and the dimensionality of spacetime. All the
spectra have a thermal profile in terms of the energy of the particle, as expected,
and bear a strong dependence on the value of the angular momentum of the black hole and
the number of extra dimensions. Next, we numerically solve the angular part of the
equation of motion to find the exact values of the spheroidal harmonics, and we
calculate the angular distribution of the flux and power spectrum; these results are also
given in section 4, and, contrary to the emission of radiation during the Schwarzschild
phase, are characterized by a strong angular variation. Finally, in the last part of
section 4 we focus on the study of the various total emissivities (the total number
of particles, energy and angular momentum emitted per unit time by the black hole in
all frequencies) and compare their behaviour for different values of the dimensionality
of spacetime and of the angular momentum of the black hole. We finish with a summary of
our results and conclusions, in Section 5.

\section{Theoretical Framework}

The gravitational background around a $(4+n)$-dimensional, rotating, uncharged black
hole was found by Myers \& Perry \cite{MP}, and in what follows we will use this
line-element to describe the spin-down phase of a small, higher-dimensional black hole
created by the collision of highly energetic particles. If we assume that the colliding
particles are restricted to propagate on an infinitely-thin 3-brane, they will have a
non-zero impact parameter only along our brane, and thus acquire only one non-zero
angular momentum parameter about an axis in the brane.
Such a higher-dimensional black hole will then be
described by the following line-element \cite{MP}
%%%%%%%%
\begin{eqnarray}
&~& \hspace*{-3cm}ds^2 = \biggl(1-\frac{\mu}{\Sigma\,r^{n-1}}\biggr) dt^2 +
\frac{2 a \mu \sin^2\theta}{\Sigma\,r^{n-1}}\,dt \, d\varphi
-\frac{\Sigma}{\Delta}\,dr^2 -\Sigma\,d\theta^2 \nonumber \\[2mm]
\hspace*{2cm}
&-& \biggl(r^2+a^2+\frac{a^2 \mu \sin^2\theta}{\Sigma\,r^{n-1}}\biggr)
\sin^2\theta\,d\varphi^2 - r^2 \cos^2\theta\, d\Omega_{n},
\label{rot-metric}
\end{eqnarray}
where
\begin{equation}
\Delta = r^2 + a^2 -\frac{\mu}{r^{n-1}}\,, \qquad
\Sigma=r^2 +a^2\,\cos^2\theta\,,
\label{Delta}
\end{equation}
%%%%%%%%%%
and $d\Omega_n$ is the line-element on a unit $n$-sphere. The mass and
angular momentum (transverse to the $r \varphi$-plane) of the black hole
are then given by
%%%%%%%%%%%
\begin{equation}
M_{BH}=\frac{(n+2) A_{n+2}}{16 \pi G}\,\mu\,,  \qquad
J=\frac{2}{n+2}\,M_{BH}\,a\,, \label{def}
\end{equation}
with $G$ being the $(4+n)$-dimensional Newton's constant, and $A_{n+2}$
the area of a $(n+2)$-dimensional unit sphere given by
%%%%%%%%%
\begin{equation}
A_{n+2}=\frac{2 \pi^{(n+3)/2}}{\Gamma[(n+3)/2]}\,.
\end{equation}
%%%%%%%%%%

In this work, we will focus on the emission of Hawking radiation \cite{Hawking},
in the form of scalar fields, from this $(4+n)$-dimensional, rotating black hole
\footnote{For some classic works on Hawking radiation in the 4-dimensional spacetime,
see \cite{classics, page, sanchez}.}. For
phenomenological reasons, we will concentrate on the emission of these fields
directly on our brane since any particle modes emitted in the bulk cannot
be detected by an observer restricted to live on the brane. We will thus
need to determine first the line-element on which the brane-localized modes
propagate, and then to solve their equation of motion on the resulting background.
The induced-on-the-brane line-element can be found by fixing the values of the
additional angular coordinates that were introduced to describe the compact extra
$n$ dimensions: by setting $\theta_i=\pi/2$, for $i=2,...,n+1$, we are led to
the 4-dimensional background
%%%%%%%%%%
\begin{equation}
\begin{split}
ds^2=\left(1-\frac{\mu}{\Sigma\,r^{n-1}}\right)dt^2&+\frac{2 a\mu\sin^2\theta}
{\Sigma\,r^{n-1}}\,dt\,d\varphi-\frac{\Sigma}{\Delta}dr^2 \\[3mm] &\hspace*{-1cm}
-\Sigma\,d\theta^2-\left(r^2+a^2+\frac{a^2\mu\sin^2\theta}{\Sigma\,r^{n-1}}\right)
\sin^2\theta\,d\varphi^2\,.
\end{split} \label{induced}
\end{equation}
%%%%%%%%

The black hole horizon is given by solving the equation $\Delta(r)=0$, which, for
$n\geq1$, leads to a unique solution given by
%%%%%%%%
\begin{equation}
r_{h}^{n+1}=\frac{\mu}{1+a_*^2}\,,
\label{horizon}
\end{equation}
%%%%%%%%
where we have defined $a_*=a/r_{h}$. In addition, while for $n=0$ and $n=1$, there is
a maximum possible value of $a$, that guarantees the existence of a real solution to
the equation $\Delta=0$ and thus of a horizon, for $n>1$ there is no fundamental
upper bound on $a$ and a horizon $r_{h}$ always exists. An upper bound can
nevertheless be imposed on the angular momentum parameter of the black hole by
demanding the creation of the black hole itself from the collision of the two
particles. The maximum value of the impact parameter between the two particles
that can lead to the creation of a black hole was found to be \cite{Harris}
%%%%%%%%%%%
\begin{equation}
b_\text{max}=2 \,\biggl[1+\biggl(\frac{n+2}{2}\biggr)^2\biggr]^{-\frac{1}{(n+1)}}
\mu^{\frac{1}{(n+1)}}\,,
\end{equation}
%%%%%%%%%%
an analytic expression that is in very good agreement with the numerical results
produced in \cite{creation}c. Then, by writing $J=b M_{BH}/2$, for the angular
momentum of the black hole, and using Eq. (\ref{horizon}) and the second of
Eq. (\ref{def}), we obtain
%%%%%%%%%%%%%
\begin{equation}
a^\text{max}_*=\frac{n+2}{2}\,.
\end{equation}
%%%%%%%%%%%%%

The equation of motion for a particle with spin 0, 1/2 or 1,  propagating in the
induced-on-the-brane gravitational background (\ref{induced}) of a higher-dimensional,
rotating black hole  was derived in \cite{Kanti, IOP1}. For scalar fields,
the field factorization
%%%%%%%%
\begin{equation}
\phi(t,r,\theta,\varphi)= e^{-i\omega t}\,e^{i m \varphi}\,R(r)\,T^{m}_{\ell}(\theta,
a \omega)\,,
\end{equation}
%%%%%%%%%
where $T^{m}_{\ell}(\theta, a \omega)$ are the so-called spheroidal harmonics
\cite{spheroidals}, leads to the following set of decoupled radial and angular
equations
%%%%%%%%%
\begin{equation}
\frac{d}{dr}\biggl(\Delta\,\frac{d R_{\omega \ell m}}{dr}\biggr)+\left(\frac{K^2}{\Delta} -\Lambda^m_{\ell}
\right)R_{\omega \ell m}=0\,, \label{radial}
\end{equation}
%%%%%%%%
\smallskip
%%%%%%%%%%
\begin{equation}
\label{spinang}
\frac{1}{\sin\theta} \frac{d}{d\theta}\left(\sin\theta\,\frac{d T^m_{\ell}(\theta, a \omega)}
{d\theta}\right) + \biggl(-\frac{m^2}{\sin^2\theta}
+a^2\omega^2\cos^2\theta + E^m_{\ell}\biggr) T^m_{\ell}(\theta, a \omega)=0\,,
\end{equation}
%%%%%%%%%
respectively. In the above, we have defined
%%%%%%%%%%%
\begin{equation}
K=(r^2+a^2)\,\omega-am\,,  \qquad
\Lambda^m_{\ell}=E^m_{\ell}+a^2\omega^2-2am\omega\,.
\end{equation}
%%%%%%%%%%
The angular eigenvalue $E^m_{\ell}$ provides a link between the angular and radial equation,
and various methods for finding its exact value will be discussed in the next section. Both
equations (\ref{radial})-(\ref{spinang}) will be solved numerically to determine the radial
and angular parts of the scalar field. The radial equation (\ref{radial}) can be solved
analytically in the asymptotic regimes of near horizon and infinity. By using the
new radial, `tortoise', coordinate
%%%%%%%%%%%%
\begin{equation}
\frac{dr^*}{dr}=\frac{r^2+a^2}{\Delta(r)}\,,
\end{equation}
%%%%%%%%%%%%,
we find the following asymptotic solution near the horizon of the black hole
%%%%%%%%%%%%%
\begin{equation}
R_{h}(r^*)=A_1\,e^{i k r^*} + A_2\,e^{-i k r^*}\,,
\label{asy-hor}
\end{equation}
%%%%%%%%%%%%%
where $A_{1,2}$ are integration constants, and $k$ is defined as
%%%%%%%%%%
\begin{equation}
k=\omega-\frac{ma}{r_{h}^2+a^2}\,.
\label{kdef}
\end{equation}
%%%%%%%%%
A boundary condition must be applied in the near-horizon regime to ensure that the solution
contains only incoming modes; this is satisfied if we set $A_1=0$. On the other hand,
for fixed $a_*$ and large $r$, the asymptotic solution at infinity takes the form
%%%%%%%%%%%%%
\begin{equation}
R_\infty(r)=B_1\,\frac{e^{i \omega r}}{r} + B_2\,\frac{e^{-i \omega r}}{r}\,,
\label{asy-inf}
\end{equation}
%%%%%%%%%%%%%
where $B_{1,2}$ are again integration constants. The asymptotic solutions (\ref{asy-hor})
and (\ref{asy-inf}) will serve as boundary conditions for our numerical analysis.

The Hawking temperature of the $(4+n)$-dimensional, rotating black hole is found to be
%%%%%%%%%
\begin{equation}
T_\text{H}=\frac{(n+1)+(n-1)a_*^2}{4\pi(1+a_*^2)r_{h}}\,,
\label{temperature}
\end{equation}
%%%%%%%%%%
and leads to the emission of thermal Hawking radiation both in the bulk and on the brane.
As mentioned earlier, here we will study the scalar field `channel' emitted on the brane.
Since we are interested in studying an evaporating black hole, the relevant
quantum state is the ``past'' Unruh vacuum $|U^{-}\rangle$ \cite{ottewill,fandt,gavin}. This
state is defined in terms of the standard ``in'' and ``up'' modes, corresponding to
unit initial particle flux at past null infinity and the past horizon, respectively
\cite{ottewill,fandt,gavin}.
At infinity, far from the black hole, the outward fluxes of energy $E$ and angular momentum $J$
can be expressed in terms of the components $T^{rt}$ and $T^{r}_{\varphi}$ of the
renormalized stress-energy tensor for a quantum scalar field in the state
$|U^{-}\rangle$ \cite{fandt}:
%%%%%%%%%%%%%%%%
\begin{eqnarray}
\frac {dE}{dt} & = &
\int _{S_{\infty }} \langle U^{-} |T^{rt} |U^{-} \rangle _{\rm {ren}}
\, r^{2} \sin \theta \, d\theta \, d\varphi \,,
\nonumber \\[1mm]
\frac {dJ}{dt} & = &
\int _{S_{\infty }} \langle U^{-} | T^{r}_{\varphi } | U^{-} \rangle _{\rm {ren}}
\, r^{2} \sin \theta \, d\theta \, d\varphi \,,
\end{eqnarray}
%%%%%%%%%%%%%%%%
where the integral is taken over the sphere at infinity.
These two components do not need renormalization for a quantum scalar field
on the metric (\ref{induced}) \cite{fandt}, thus simplifying the computation.
Our boundary conditions, (\ref{asy-hor}) and (\ref{asy-inf}), correspond
to the ``in'' modes \cite{ottewill,gavin}, and in terms of these the relevant quantities can be
written as\footnote{Wronskian relations
between the ``in'' and ``up'' modes allow us to easily change our basis; in terms of the
``up'' modes, the quantity $|{\cal A}_{\ell,m}|^2 \equiv|{\cal A}^{in}_{\ell,m}|^2$ in
Eqs. (\ref{power1})-(\ref{ang-mom1}) is replaced by
$(\omega/k)\,|{B}^{-}_{\omega \ell m}|^2$ in the notation of \cite{ottewill}.}
%%%%%%%%%%%%%%%
\begin{eqnarray}
\langle U^{-}| T^{rt} | U^{-} \rangle _{\rm {ren}} & = &
\frac {1}{4\pi ^{2}r^{2}} \sum _{\ell ,m}
\int _{\omega =0 }^{\infty }
\frac {\omega \, d\omega }{\exp\left[k/T_\text{H}\right] - 1}
|{\cal A}_{\ell,m}|^2 \,
|T^m_{\ell}(\theta, a \omega) |^{2}\,,
\label{power1}
\\
\langle U^{-}| T^{r}_{\varphi } | U^{-} \rangle _{\rm {ren}} & = &
\frac {1}{4\pi ^{2}r^{2}} \sum _{\ell ,m}
\int _{\omega =0 }^{\infty }
\frac {m \, d\omega }{\exp\left[k/T_\text{H}\right] - 1}
|{\cal A}_{\ell,m}|^2 \,
|T^m_{\ell}(\theta, a \omega) |^{2}\,.
\label{ang-mom1}
\end{eqnarray}
%%%%%%%%%%%%%%%
In the above expressions,
$|{\cal A}_{\ell,m}|^2$ is the absorption (or transmission) probability for a
scalar particle propagating in the brane background (\ref{induced}),
and $k$ is given by (\ref{kdef}). Multiplying
these expressions by $r^{2}\sin \theta $ and integrating over the sphere at infinity
gives the total outgoing fluxes of energy and angular momentum:
\begin{eqnarray}
\frac {dE}{dt}   &=&
\frac {1}{2\pi} \sum _{\ell ,m}
\int _{\omega =0 }^{\infty }
\frac {\omega \, d\omega }{\exp\left[k/T_\text{H}\right] - 1}
|{\cal A}_{\ell,m}|^2\,, \\[2mm]
\frac {dJ}{dt}  &=&
\frac {1}{2\pi } \sum _{\ell ,m}
\int _{\omega =0 }^{\infty }
\frac {m \, d\omega }{\exp\left[k/T_\text{H}\right] - 1}
|{\cal A}_{\ell,m}|^2 \,,
\end{eqnarray}
where we have used the fact that the spheroidal harmonics $T^m_{\ell}(\theta , a\omega )$
are normalized
according to
\begin{equation}
\int _{\theta =0}^{\pi } d\theta \, \sin \theta\, | T^m_{\ell } (\theta, a \omega) |^{2}
=1 .
\label{Tnorm}
\end{equation}
%%%%%%%%%%%
The absorption probability $|{\cal A}_{\ell,m}|^2$ has an explicit dependence not only
on the angular momentum numbers $(\ell, m)$, as denoted, but also on the energy $\omega$
of the emitted particle and the number of extra dimensions $n$. Therefore, its presence
in the expressions of the emission rates modifies significantly the blackbody profile
of the spectrum, especially in the low- and intermediate-energy regime. Its exact
form will be found by solving numerically
the radial equation (\ref{radial}) and determining the amplitudes of the incoming and
outgoing modes at infinity, according to the asymptotic solution (\ref{asy-inf}). Then,
the absorption probability may be written as
\begin{equation}
|{\cal A}_{\ell,m}|^2=1-|{\cal R}_{\ell,m}|^2=1-\biggl|\frac{B_1}{B_2}\biggr|^2\,.
\label{abs}
\end{equation}

The first part of this work will focus on the computation of the differential
emission rates of particles (flux spectrum), energy (power spectrum) and angular
momentum, integrated over all angles $\theta $. These are given by
%%%%%%%%%%%%%%%%%
\begin{eqnarray}
\frac {d^{2}N}{dt  d\omega}  & = &
\frac {1}{2\pi} \sum _{\ell ,m}
\frac {1}{\exp\left[k/T_\text{H}\right] - 1}
|{\cal A}_{\ell,m}|^2\,,
\label{flux} \\[1mm]
\frac {d^{2}E}{dt  d\omega }  &  = &
\frac {1}{2\pi} \sum _{\ell ,m}
\frac {\omega }{\exp\left[k/T_\text{H}\right] - 1}
|{\cal A}_{\ell,m}|^2\,,
\label{power} \\[1mm]
\frac {d^{2}J}{dt  d\omega}  & = &
\frac {1}{2\pi } \sum _{\ell ,m}
\frac {m}{\exp\left[k/T_\text{H}\right] - 1}
|{\cal A}_{\ell,m}|^2\,. \label{ang-mom}
\end{eqnarray}
%%%%%%%%%%%%%%
By using the exact solutions of the angular equation (\ref{spinang}), we will then study
the angular distribution of the differential fluxes of particles and energy due to the
axial symmetry of the gravitational background -- a feature
that will be a distinct signature of emission during the spin-down phase:
%%%%%%%%%%%%
\begin{equation}
\frac {d^{3}N}{d(\cos \theta )dt  d\omega}   =
\frac {1}{2\pi} \sum _{\ell ,m}
\frac {1}{\exp\left[k/T_\text{H}\right] - 1}
|{\cal A}_{\ell,m}|^2 |T^m_{\ell}(\theta, a \omega) |^{2} \,,
\label{flux-ang}
\end{equation}
\begin{equation}
\frac {d^{3}E}{d(\cos \theta )dt  d\omega }   =
\frac {1}{2\pi} \sum _{\ell ,m}
\frac {\omega }{\exp\left[k/T_\text{H}\right] - 1}
|{\cal A}_{\ell,m}|^2 |T^m_{\ell}(\theta, a \omega) |^{2} \,.
\label{pow-ang}
\end{equation}
In both cases, the various spectra will be given as a function of the energy of the
emitted mode $\omega$, but their dependence on the angular momentum parameter of the
black hole and the dimensionality of spacetime will be also studied in detail.

\section{Numerical Analysis}
\label{sec:numerics}

For the purpose of the analysis presented in this work, we will need to solve
both the angular (\ref{spinang}) and the radial (\ref{radial}) part of the
scalar equation. The angular equation will be studied first as the eigenvalue
$E^m_{\ell}$ is required before we can integrate the radial equation.
The spheroidal harmonics, which are the solutions of Eq. (\ref{spinang}),
have been extensively studied in the literature \cite{spheroidals}, therefore,
here we will only briefly outline our method and some of their key properties.

Before describing the numerical analysis performed in this work though, we would
like to note that if one is interested in solving the radial equation only and
studying the spectrum integrated over the angle $\theta$, then it is sufficient
to find only the eigenvalues $E^m_{\ell}$, while the eigenfunctions
$T^m_{\ell}(\theta , a\omega )$ are not required.
The eigenvalues $E^m_{\ell}$ are functions of $a\omega $, and an analytic form
exists only in the limit of small $a\omega $. To derive results for the
emission of Hawking radiation valid for arbitrarily large energy of the emitted
particle and angular momentum of the black hole, one needs to find numerically the
exact value of $E^m_{\ell}$. This can be done, for instance, by using the so-called
continuation method \cite{Wasserstrom}.  The continuation method is a generalization
of perturbation theory which is applicable when the changes in the initial Hamiltonian
(for which the eigenvalues are known) are not necessarily small. It involves writing
the $T^m_{\ell}(\theta,a\omega)$ functions in the basis of the $\theta$-parts of the
spherical harmonics, $S_{\ell}^m(\theta)$, which in conjunction with the angular
equation and techniques from the perturbation theory leads to a differential
equation for the eigenvalue $E^m_{\ell}$. This is, then, solved numerically, by
using appropriate initial conditions, to derive the eigenvalue for any $\ell$ and
$m$ and for any value of $a\omega$. This method was outlined in \cite{Harris}
and used in \cite{HK2} for the derivation of the exact power spectrum for the emission
of Hawking radiation, in the form of scalar fields, from a 5-dimensional black hole
on the brane.

As mentioned above, in this work we also  compute the exact angular
distributions of the fluxes of particles and energy, therefore we require the
values of the spheroidal harmonics $T^m_{\ell}(\theta, a \omega)$ themselves.
For this, we follow a modified version of the method presented in section 17.4
of \cite{NR}. Given the symmetry
in the angular equation (\ref{spinang}) under the replacement of $m$ by $-m$, it
is sufficient to consider only those solutions for which $m \ge 0$.
It is also convenient to change the independent variable to $\eta = \cos \theta $ and
define a new dependent variable $y_{\ell m \omega}$ by
\begin{equation}
y_{\ell m \omega} = \left( 1- \eta ^{2} \right) ^{-\frac {m}{2}} T^m_{\ell}(\theta,
a \omega),
\end{equation}
in terms of which the angular equation (\ref{spinang}) becomes
\begin{eqnarray}
\left( 1- \eta ^{2} \right) \frac {d^{2}y_{\ell m \omega } }{d\eta ^{2}} -
2\left( m+1 \right) \eta \frac {dy_{\ell m \omega } }{d\eta }
+ \left( E^m_{\ell} -m \left( m+1 \right) + a^{2} \omega ^{2} \eta ^{2}
 \right) y_{\ell m \omega } =0\,.
 \label{angalt}
\end{eqnarray}
The above equation must be solved on the interval $[-1,1]$, with the boundary conditions
that $y_{\ell m \omega}$ is regular and non-zero at both end-points.
We used a shooting method \cite{NR} to solve this boundary value problem.
For spin 0 particles, the number of zeros of $y_{\ell m \omega}$ in the interval
$[-1,1]$ is given by $\ell -m$, and the spheroidal harmonics are either odd or
even functions of $\eta $. This symmetry can be used to integrate Eq. (\ref{angalt})
over the half-interval $[-1,0]$, using appropriate boundary conditions at $\eta =0$.
However, we have chosen not to take this path, but instead to integrate over the
full interval $[-1,1]$ as this method will more readily generalize to higher spins
\cite{CKW}, when the spin-weighted spheroidal harmonics do not have this symmetry property.

%%%%%%%%%%%%%%%
\begin{figure}[t]
\begin{center}
\includegraphics[height=6cm,clip]{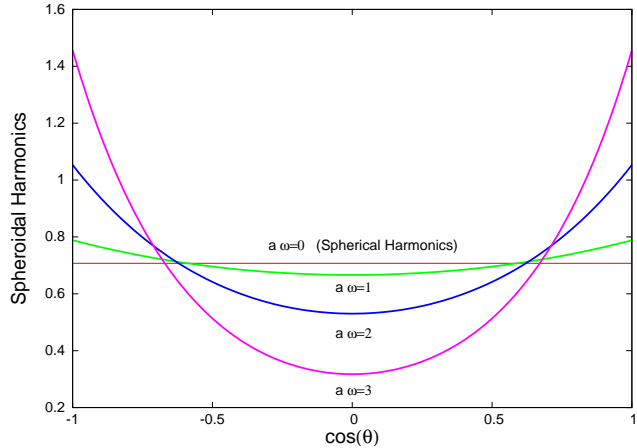}
\caption{Spheroidal harmonics for a scalar field with $\ell=m=0$, and for
$a \omega=(0,1,2,3)$. }
\label{spheroidal}
\end{center}
\end{figure}
%%%%%%%%%%%%%%%

The end-points, $\eta =\pm 1$, are regular singular points of Eq. (\ref{angalt}). By
using the Frobenius method, we find two linearly independent solutions near these
end-points, one of which is regular and the other irregular. We begin the numerical
integration at $\eta =-1+\delta\eta $, with $\delta\eta =10^{-7}$, using the Frobenius
series expansion to impose our initial condition for a regular function $y$. We then
integrate out to $\eta =1$, and demand that the solution be also regular at $\eta =1$.
For a general initial value of $E^m_{\ell}$, the numerical solution, close to $\eta =1$,
will be a linear combination of the regular solution we are seeking and the irregular
solution. We define a function $F$, which depends on $E^m_{\ell}$, as follows:
\begin{equation}
F(E^m_\ell)=
\left. \frac {1}{y_{\rm {reg}}} \frac {dy_{\rm {reg}}}{d\eta } \right|_{\eta =1}
- \left. \frac {1}{y} \frac {dy}{d\eta } \right|_{\eta =1}\,,
\end{equation}
where $y$ is the numerical solution we have found by integration,
and $y_{\rm {reg}}$ is the solution
regular at $\eta =1$. The idea is therefore to find the value of $E^m_{\ell}$ such that
$F(E^m_{\ell})=0$. Due to the irregular solution, the integration cannot be extended all
the way to $\eta =1$, but instead we impose the boundary condition at $\eta = 1-\delta\eta$, again using a Frobenius series expansion for the regular solution $y_{\rm {reg}}$.

The above method can give us the values of both the angular eigenvalues $E^m_{\ell}$
and the spheroidal harmonics $T^m_{\ell}(\theta, a \omega)$. In the case of the eigenvalues,
we have used both the continuation and the shooting method, and we have found an
excellent agreement between the values obtained. For the spheroidal eigenfunctions,
the results found by using the shooting method\,\footnote{Further details of this
numerical method, generalized for all spins, can be found in \cite{marc} -- an
alternative method of finding the spheroidal harmonics, once the eigenvalues are
known, is presented in \cite{leaver}.} were finally normalized according to
Eq. (\ref{Tnorm}). In Fig. \ref{spheroidal}, as an illustrative example,
we depict the derived spheroidal
harmonics for the scalar mode with $\ell=m=0$, and for the values $a \omega=0,1,2$
and 3 -- we remind the reader that, for $a \omega =0$, the spheroidal harmonics
reduce to the spherical ones. In previous works \cite{IOP1,IOP2}, the spherical
harmonics were used as an approximation to the exact spheroidal ones. From
Fig. \ref{spheroidal}, we can clearly see that, as long as $a \omega \ll 1$,
this is indeed a valid approximation. However, as $a \omega $ increases, this
approximation becomes increasingly poor. In addition, the difference between the
spheroidal harmonics and the spherical ones is strongly mode-dependent, being more
significant for the low ($\ell, m$) modes and less so for the high ($\ell, m$) modes.
For an accurate analysis therefore, leading to the exact angular distribution of particle
and energy fluxes, valid at all energy and angular momentum regimes, the use of
the exact spheroidal harmonics is imperative.

Having derived the values of the angular eigenvalues  $E_\ell^m$, the integration of
the radial differential equation (\ref{radial}) can now proceed. We start the integration
at the horizon of the black hole and integrate outwards towards infinity. Comparing
our numerical results for the radial solution with the asymptotic solution at infinity
(\ref{asy-inf}), we determine the integration constants $B_1$ and $B_2$. The absorption
probability for scalar fields can then be derived from the relation (\ref{abs}). The
use of the exact value of the angular eigenvalue $E_\ell^m$, instead of an approximate
analytical formula, guarantees the validity of the solution of the radial equation, too,
for arbitrary energy of the emitted particle and angular momentum of the black hole.

\section{Numerical Results}

In this section, we present exact numerical results for the flux, power and angular momentum
spectra for the emission of scalar particles on the brane from a higher-dimensional, rotating
black hole. The dependence of the various spectra on the two fundamental parameters --
the dimensionality of spacetime $n$ and the angular momentum parameter $a_*$ -- will be
studied in detail, and in the case of the first two types of spectra (flux and power),
the angular distribution of the emitted particles and energy will also be examined. We will
finally study the corresponding total emissivities of the black hole on the brane. These
follow by integrating the various emission spectra over the whole frequency regime, and can provide information on the dependence of the total number of particles, energy and angular
momentum, emitted by the black hole per unit time, on $n$ and $a_*$, as well as on their
relative behaviour as the values of $n$ and $a_*$ vary.

\subsection{Flux emission spectra}

\begin{figure}[t]
\begin{center}
\thicklines
\mbox{\psfrag{x}[][][0.8]{$\omega\,r_\text{h}$}
\psfrag{y}[][][0.8]{${r_\text{h}^2\,d^2N/dt\,d\omega}$}
\includegraphics[height=5.6cm,clip]{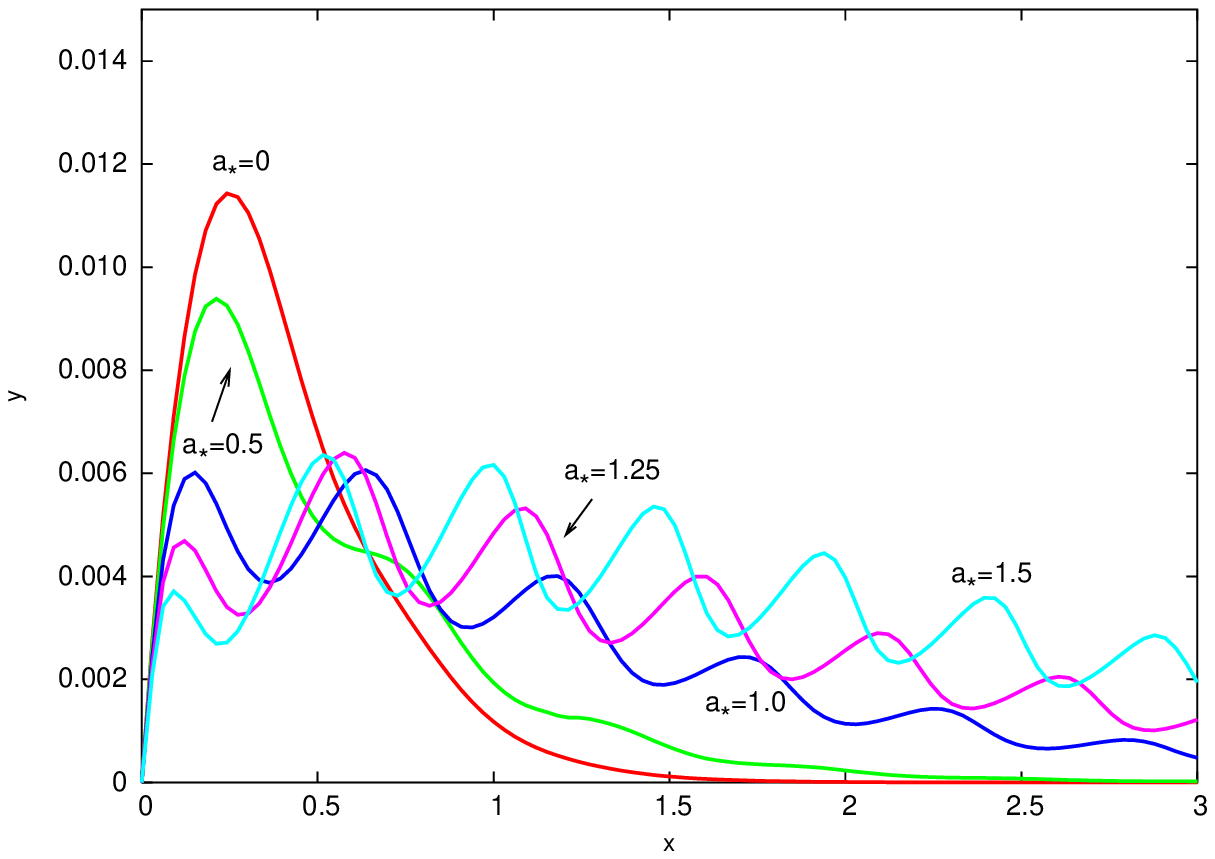}}%\hspace*{0.1cm}{
{\psfrag{x}[][][0.8]{$\omega\,r_\text{h}$}
\psfrag{y}[][][0.8]{$r_\text{h}^2\,d^2N/dt\,d\omega$}
\includegraphics[height=5.6cm,clip]{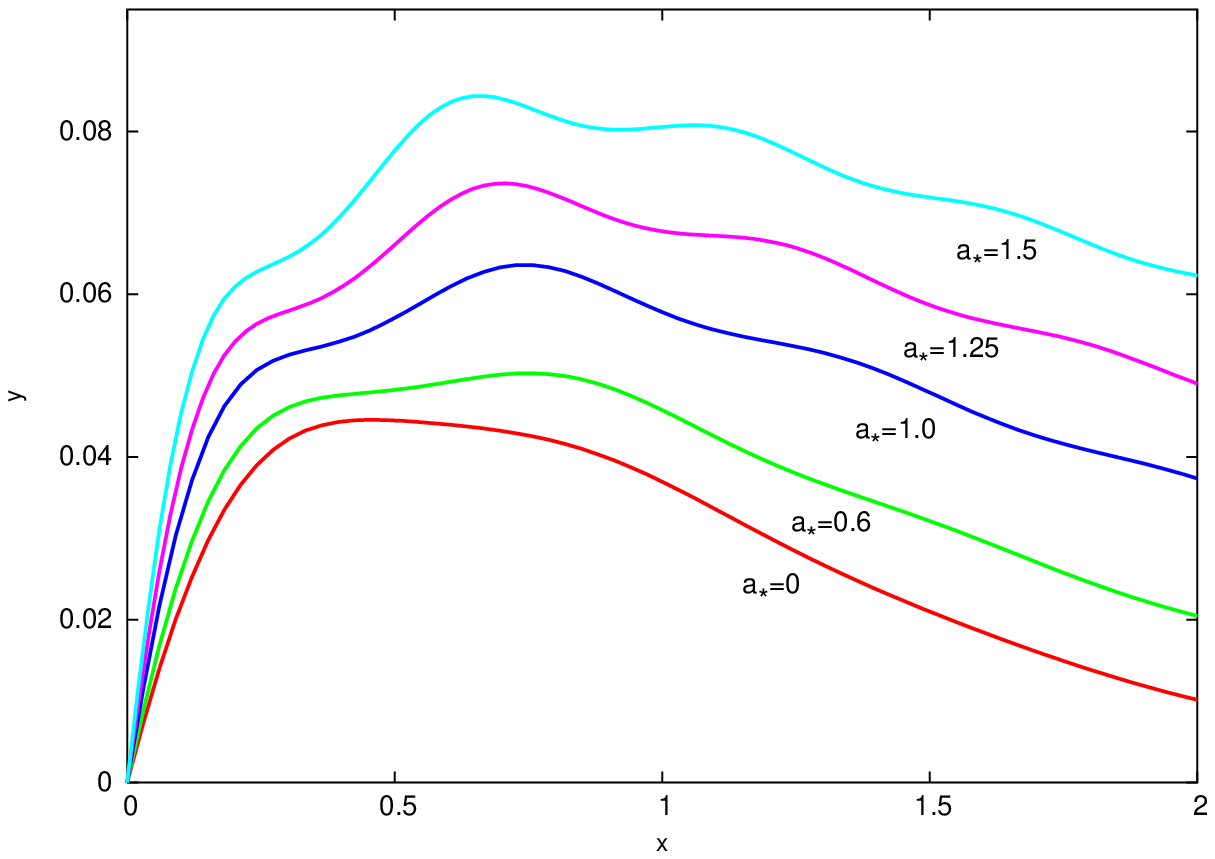}}
%\scalebox{0.8}{\rotatebox{0}{\includegraphics[width=\textwidth]{nefiopfig2.eps}}}
\caption{Flux emission spectra for scalar particles on the brane from a rotating black hole,
for (a) $n=1$, and (b) $n=4$, and various values of $a_*$.\hspace*{1.5cm}}
\label{s0n1-4flux}
\end{center}
\end{figure}

We start by presenting our numerical results for the number of scalar particles emitted
by the black hole on the brane per unit time and unit frequency. Figures \ref{s0n1-4flux}(a,b)
depict the flux emission rate, Eq. (\ref{flux}), as a function of the frequency of the emitted
particle in dimensionless units\footnote{Throughout this analysis, we will be assuming that
the horizon radius $r_h$ of the black hole remains fixed as $n$ and $a_*$ varies.},
$\omega r_h$, for various values of the angular momentum parameter $a_*$ of the
black hole. The behaviour of the flux depends strongly also on the dimensionality of spacetime, and, for that reason, Figs. \ref{s0n1-4flux}(a) and (b) present results for two indicative
cases, $n=1$ and $n=4$, respectively.

\smallskip
%%%%%%%%%%%%%%%
%%%%%%%%%%%%%%
%

In the case of a 5-dimensional black hole, the particle flux, shown in Fig. \ref{s0n1-4flux}(a),
is characterized, for small values of $a_{*}$, by a strong peak at low values of $\omega r_{h}$;
this clearly favours the emission of low-energy quanta. As the angular momentum parameter
increases, however, this peaked Gaussian curve gives its place to a broad one,
where high-energy particles become almost as equally likely to be emitted as the low- and
intermediate-energy ones. This feature will be of prime importance in the determination
of the power spectrum, to be studied in the next subsection. For higher
values of $a_*$, oscillations become a common feature of the spectrum; these are due to
the higher partial waves gradually coming into dominance as the energy increases.
In Fig. \ref{s0n1-4flux}(b), we depict the flux spectrum of an 8-dimensional rotating black
hole. The spectrum shares common characteristics with the one obtained in the 5-dimensional
case, with the curve becoming broader and the tail dying away much more slowly, as $a_{*}$
increases. However, there is now a clear enhancement in the number of particles emitted
by the black hole per unit time over the whole energy regime, as $a_*$ increases, and the
oscillations, although still present, are less apparent here.

\begin{figure}[bh]
\begin{center}
\psfrag{x}[][][0.8]{$\omega\,r_\text{h}$}
\psfrag{y}[][][0.8]{$r_\text{h}^2\,d^2N/dt\,d\omega$}
\psfig{file=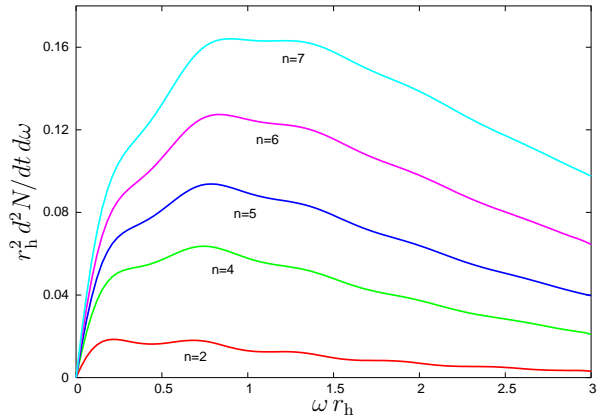, height=5.6cm}
%\includegraphics[height=5.6cm,clip]{fluxs0a1.eps}
%\scalebox{0.8}{\rotatebox{0}{\includegraphics[width=\textwidth]{nefiopfig2.eps}}}
\caption{Flux emission spectra for scalar particles on the brane from a rotating black hole,
with $a_*=1$, for various values of $n$.\hspace*{1.5cm}}
\label{s0a1flux}
\end{center}
\end{figure}

Comparing the vertical axes of Figs. \ref{s0n1-4flux}(a) and (b), one easily observes
the almost one order-of-magnitude enhancement in the flux of particles, as
we go from the $n=1$ to the $n=4$ case. This enhancement was a characteristic feature
of the flux spectrum in the case of a non-rotating black hole \cite{HK1}, and -- as we
see in Fig. \ref{s0a1flux} -- it characterizes also the spectrum of a rotating black hole.
Figure \ref{s0a1flux} depicts the particle emission rate for a rotating black hole with
a fixed angular momentum parameter, $a_*=1$, living in spacetimes with different dimensionalities. The particle flux broadly increases, and peaks for higher values of
$\omega r_{h}$, as $n$ increases, clearly leading to a higher total number of particles
emitted per unit time, as we will also see in subsection 3.4.

\medskip
%%%%%%%%%%%%%%%
%%%%%%%%%%%%%%

%\smallskip
%%%%%%%%%%%%%%%
\begin{figure}[t]
\begin{center}
\begin{tabular}{c} \hspace*{-0.4cm}
\epsfig{file=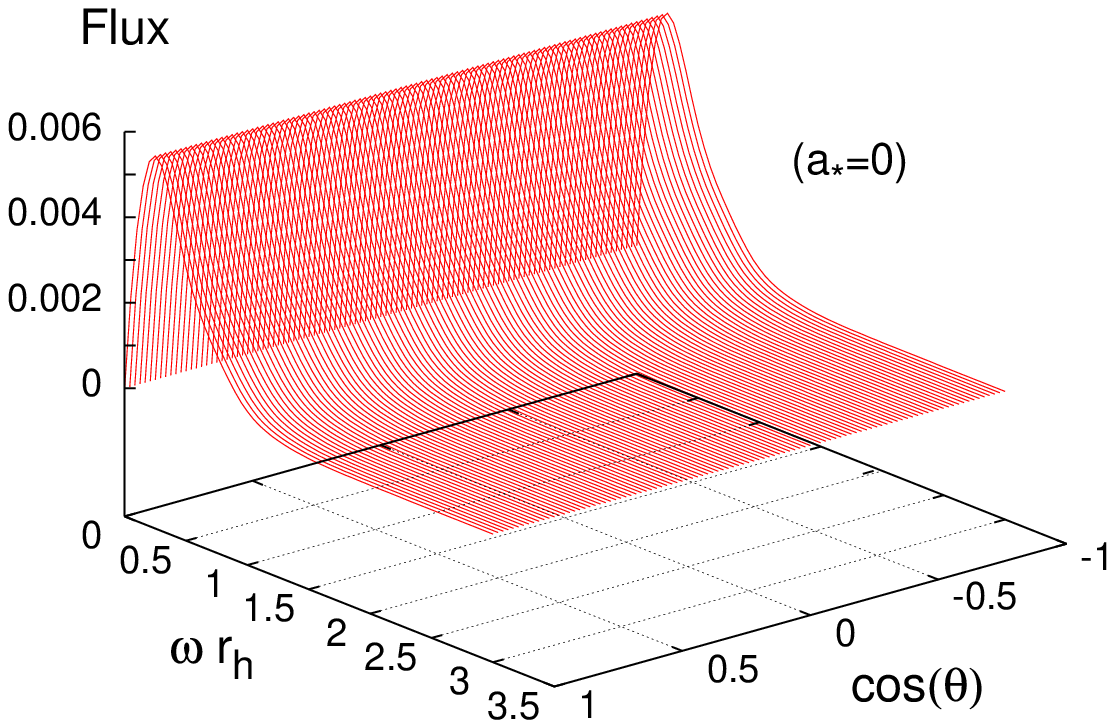, height=4.15cm}\hspace*{-1.15cm}
\epsfig{file=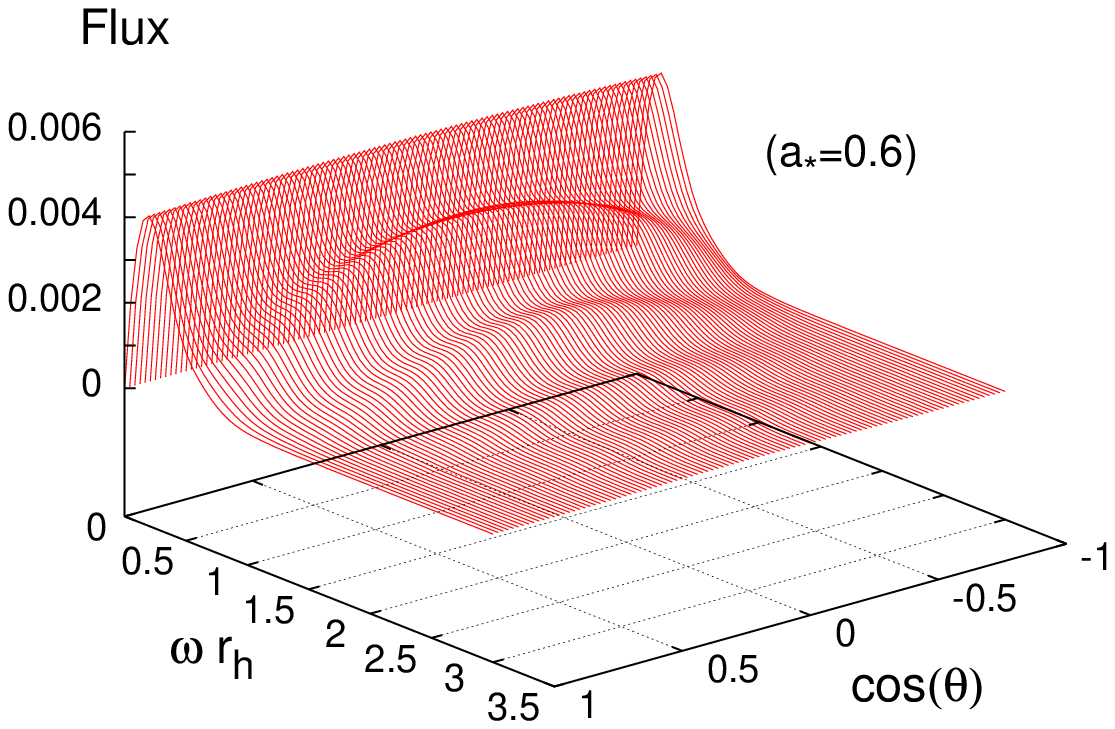, height=4.15cm}\hspace*{-1.15cm}{
\epsfig{file=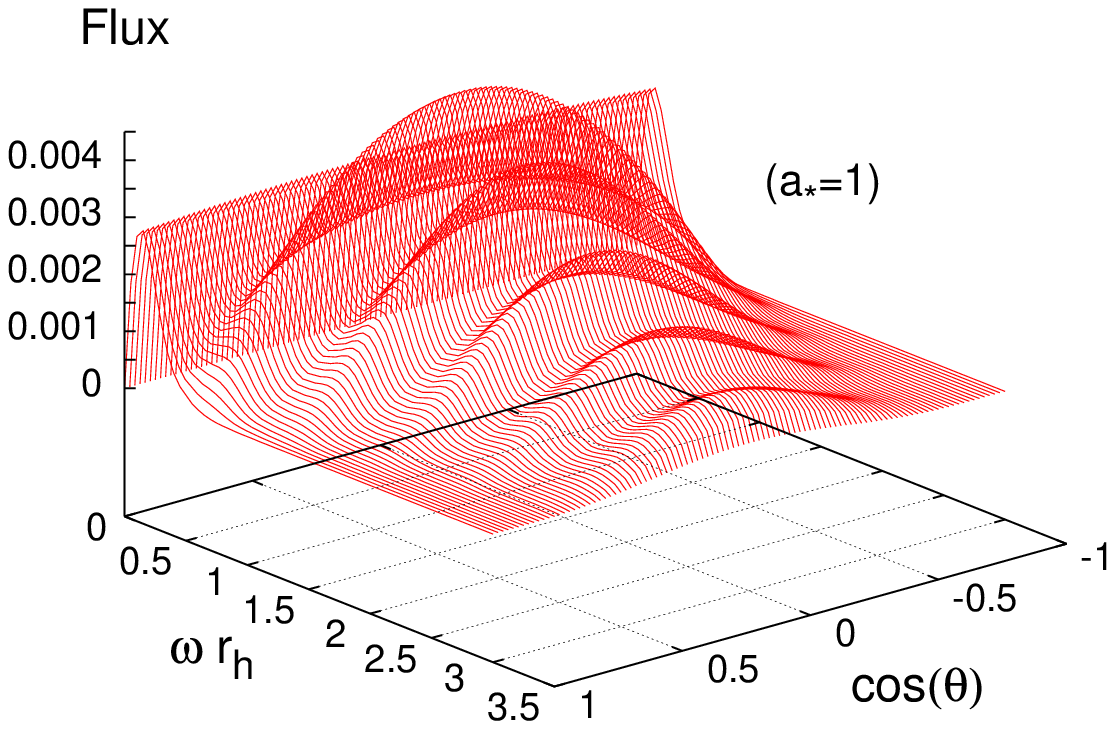, height=4.15cm}}\end{tabular}
\caption{Angular distribution of the flux spectra for scalar emission on the brane
from a rotating black hole, for $n=1$ and $a_*=(0,0.6,1)$.}
\label{s0n1f-ang}
\end{center}
\end{figure}
%%%%%%%%%%%%%%
%%%%%%%%%%%%%%%
\begin{figure}[t]
\begin{center}
\begin{tabular}{c} \hspace*{-0.4cm}
\epsfig{file=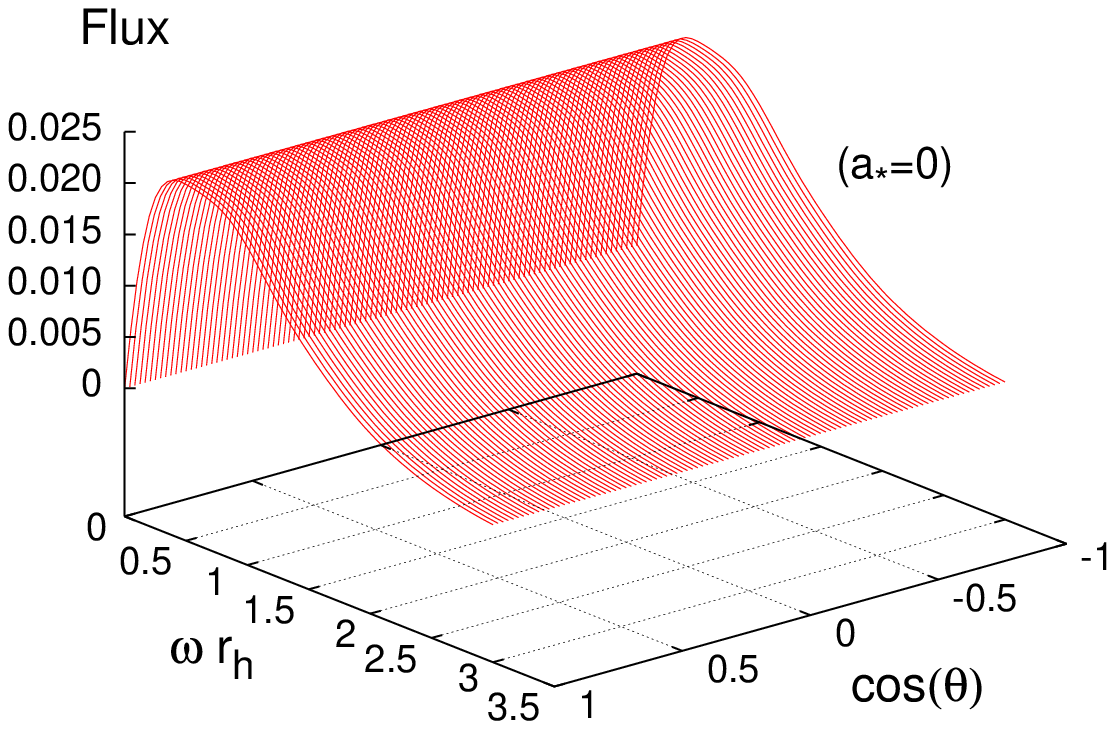, height=4.15cm}\hspace*{-1.15cm}
\epsfig{file=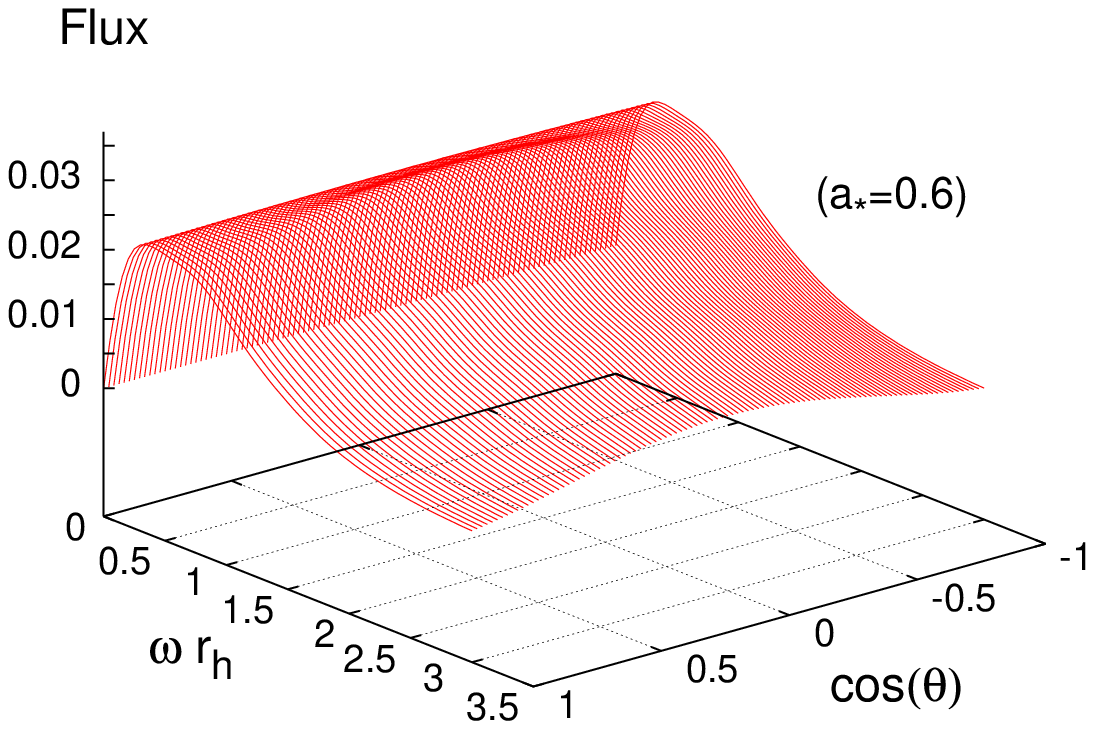, height=4.15cm}\hspace*{-1.15cm}{
\epsfig{file=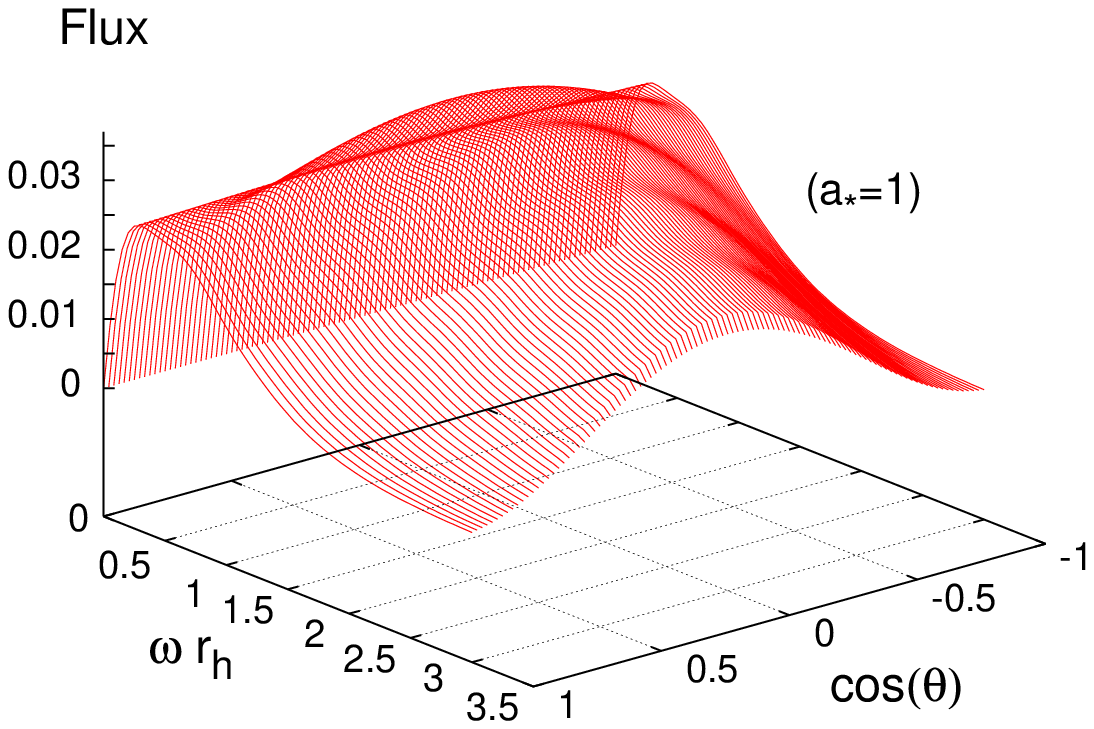, height=4.15cm}}\end{tabular}
\caption{Angular distribution of the flux spectra for scalar emission on the brane
from a rotating black hole, for $n=4$ and $a_*=(0,0.6,1)$.}
\label{s0n4f-ang}
\end{center}
\end{figure}
%%%%%%%%%%%%%%
%\smallskip

Also of great importance is the angular distribution of the emitted particles. Where\-as,
in the case of a non-rotating black hole, we expect a uniform distribution of particles,
in the case of a rotating black hole, we anticipate a distinctly different angular
distribution pattern. An angular variation in the flux and power spectra would be a clear,
characteristic signature for emission during the {\it spin-down phase} in the life
of a black hole. Figures \ref{s0n1f-ang} and \ref{s0n4f-ang} depict the flux emission
spectra on the brane from a rotating black hole with $n=1$ and $n=4$, respectively, and
for $a_*=(0,0.6,1)$, as a function of the energy of the emitted particle and the value
of $\cos\theta$ of the azimuthal angle. As expected, for vanishing $a_*$, the spectrum
shows no angular variation, independently of the value of $n$. As the angular momentum parameter
increases, the emitted modes are starting to concentrate on a region around the equator
($\theta = \pi /2$), however, their angular distribution depends, at the same time,
also on their energy: even for non-vanishing $a_*$, at low energies
the spectrum is dominated by a spherically-symmetric distribution, and shows no
dependence on the angle $\theta$; only when the energy exceeds a certain value does the
non-spherically-symmetric nature of the emission become significant.
% important enough to introduce an angular
%variation in the spectrum.
%As $a_{*}$ increases, the oscillations make their
%appearance, more prominently in the $n=1$ case, and less so in the $n=4$, in agreement
%with the previous results.
%For $n=4$, the tail of the flux dies off much more slowly, and
%we can see the anticipated enhancement in the high-energy regime.
In general, for
fixed $a_*$, as $n$ increases, the non-spherically-symmetric behaviour becomes dominant
sooner, i.e. at lower energies, and there is an increasing contribution from angles
away from the equator~\footnote{We should note here that after the end of our
analysis, the angular distribution of both the particle and energy flux was integrated
over $\cos \theta $ to check that this produces the original particle flux and power
spectrum. The two results agreed to an accuracy of $10^{-6}$.}.

\smallskip
%%%%%%%%%%%%%%%
\begin{figure}[t]
\begin{center}
\mbox{\psfrag{x}[][][0.8]{$\omega\,r_\text{h}$}
\psfrag{y}[][][0.8]{$r_\text{h}\,d^2E/dt\,d\omega$}
\includegraphics[height=5.6cm,clip]{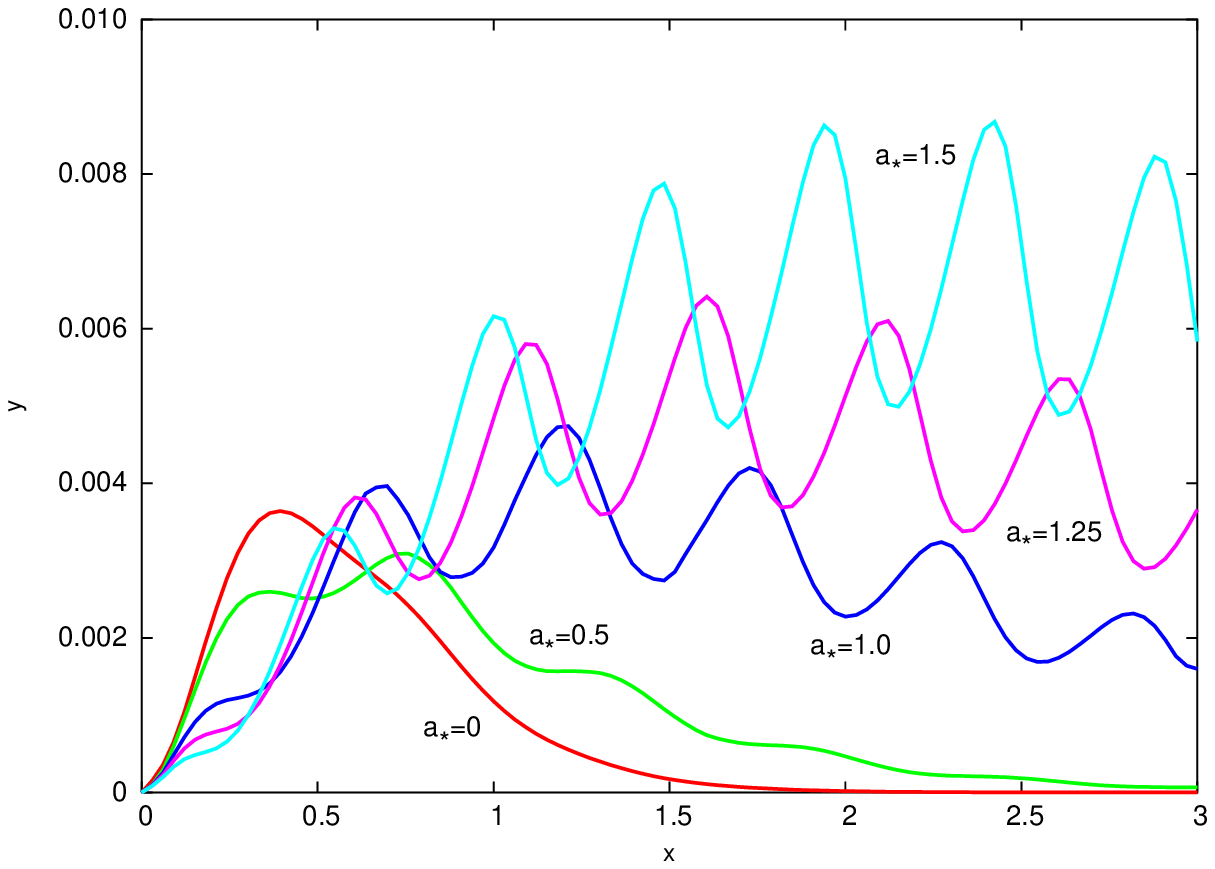}}%\hspace*{0.1cm}{
{\psfrag{x}[][][0.8]{$\omega\,r_\text{h}$}
\psfrag{y}[][][0.8]{$r_\text{h}\,d^2E/dt\,d\omega$}
\includegraphics[height=5.6cm,clip]{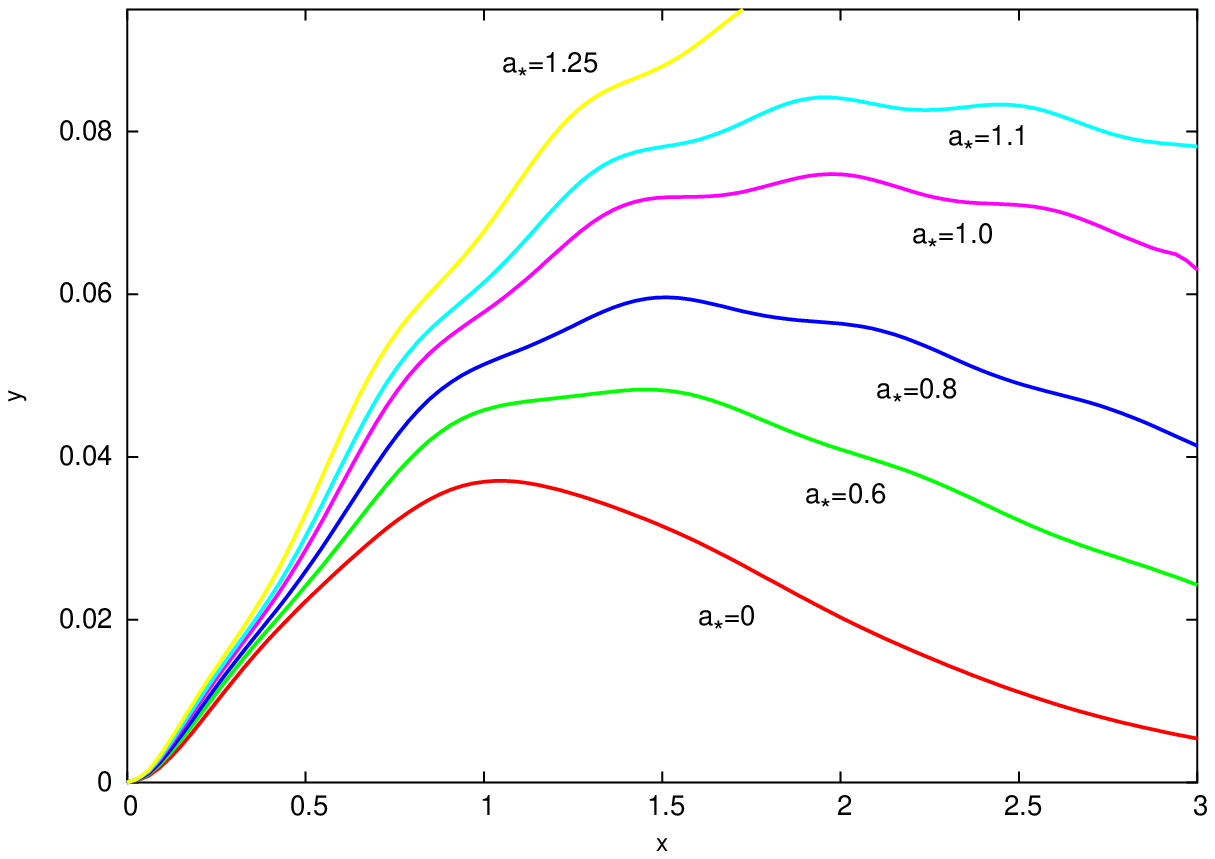}}
%\scalebox{0.8}{\rotatebox{0}{\includegraphics[width=\textwidth]{nefiopfig2.eps}}}
\caption{Power emission spectra for scalar particles on the brane from a rotating black hole,
for (a) $n=1$, and (b) $n=4$, and various values of $a_*$.\hspace*{1.5cm}}
\label{s0n1power}
\end{center}
\end{figure}
%%%%%%%%%%%%%%

\subsection{Power emission spectra}

We now move to the power spectrum that describes the energy emitted by the black hole on the
brane in the form of scalar particles per unit time and unit frequency. As in the previous
subsection, in Figs. \ref{s0n1power}(a,b), we display the energy emission spectra, for
$n=1$ and $n=4$ respectively, as a function of the frequency of the emitted particle
$\omega r_h$ and for various values of the angular momentum parameter $a_*$ of the black hole.

Since the energy emission rate follows by multiplying the particle emission rate by the frequency of the emitted particle, the numerical results shown in Figs. \ref{s0n1power}(a,b)
can be easily justified by using the ones for the flux spectrum derived in the previous subsection. In the case of a 5-dimensional black hole, we found that, for small
values of $a_*$, the black hole prefers to emit low-energy quanta, while, for higher
values of $a_*$, the emission of low-energy particles is suppressed while that of
intermediate- and high-energy particles becomes equally likely. Therefore, as the angular
momentum parameter increases, we expect the energy emitted by the black hole per unit time
to be suppressed in the low-energy regime, and to significantly increase in the
high-energy regime -- this is exactly the behaviour depicted
%\,\footnote{In Ref.
%\cite{HK2}, a similar graph was produced for a 5-dimensional rotating black hole with
%a slightly different set of parameters - although the two figures have been produced by
%using two different numerical codes, the agreement in the general behaviour is remarkable.}
in Fig. \ref{s0n1power}(a).

In the case of an 8-dimensional black hole, however, the flux spectrum [see Fig.
\ref{s0n1-4flux}(b)] shows no suppression even at the low-energy regime. As a result, the
energy emission rate is uniformly enhanced, as $a_*$ increases, over the whole frequency
regime. This behaviour, shown in Fig. \ref{s0n1power}(b), seems to be in disagreement with
the one derived in Ref. \cite{IOP2}, where the suppression at the low-energy regime in the
power spectrum persists for all values -- low or high -- of the dimensionality of spacetime.

%%%%%%%%%%%%%%%
%%%%%%%%%%%%%%%
\begin{figure}
\begin{center}
\psfrag{x}[][][0.8]{$\omega\,r_\text{h}$}
\psfrag{y}[][][0.8]{$r_\text{h}\,d^2E/dt\,d\omega$}
\includegraphics[height=5.6cm,clip]{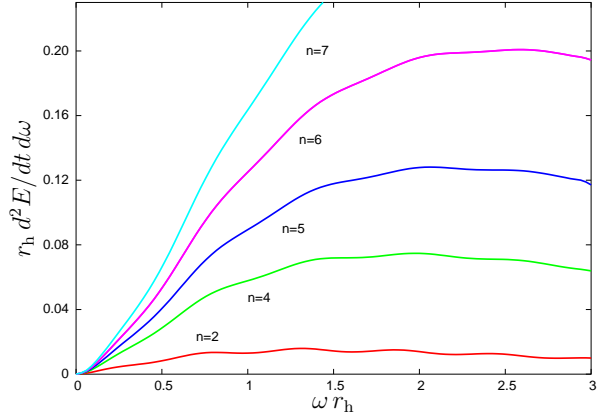}
%\scalebox{0.8}{\rotatebox{0}{\includegraphics[width=\textwidth]{nefiopfig2.eps}}}
\caption{Power emission spectra for scalar particles on the brane from a rotating black hole,
with $a_*=1$, for various values of $n$.\hspace*{1.5cm}}
\label{s0a1fp}
\end{center}
\end{figure}
%%%%%%%%%%%%%%

Other features characterizing the flux spectra are also transferred almost intact to the power
spectra. The oscillations make again their appearance, as the angular momentum increases, being
quite prominent in the $n=1$ case and smoothed out for $n=4$. In both cases,
the emission curves peak at higher values of $\omega r_{h}$ for larger values of $a_{*}$,
signifying a clear enhancement of the total emission rate, as we will see in detail in
Section 3.4. By comparing again the vertical axes of Figs. \ref{s0n1power}(a) and (b),
an enhancement of more than one order-of-magnitude in the energy emission rate of the black
hole is easily observed, as we go from $n=1$ to $n=4$. The enhancement in the power spectrum,
as the dimensionality of spacetime increases, is clearly associated with the one
found for the flux spectrum in the previous subsection, and is shown in detail in
Fig. \ref{s0a1fp}, for fixed $a_*$ and various values of $n$. A similar, strong enhancement
in the power spectrum was also found in the case of a non-rotating black hole in \cite{HK1},
and independently confirmed by other subsequent works \cite{Doran, Jung-charge, BGK}. In
the case of higher-dimensional, rotating black holes, the only other available results in the
literature are the ones presented in \cite{IOP2}; however, the figures in that paper show only an
infinitesimal enhancement of the power spectrum as $n$ increases, and curiously
enough the enhancement is also missing from the results in \cite{IOP2}
for the non-rotating, limiting
case with $a_*=0$.

\begin{figure}[ht]
\begin{center}
\begin{tabular}{c} \hspace*{-0.4cm}
\epsfig{file=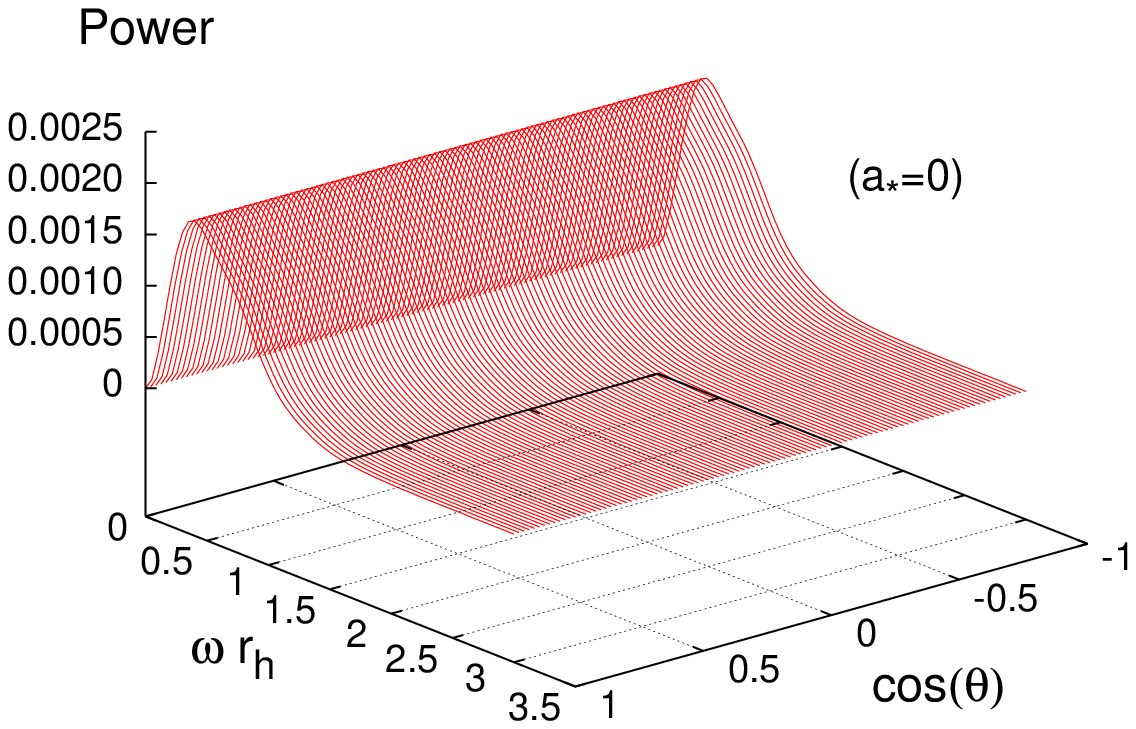, height=4.15cm}\hspace*{-1.15cm}
\epsfig{file=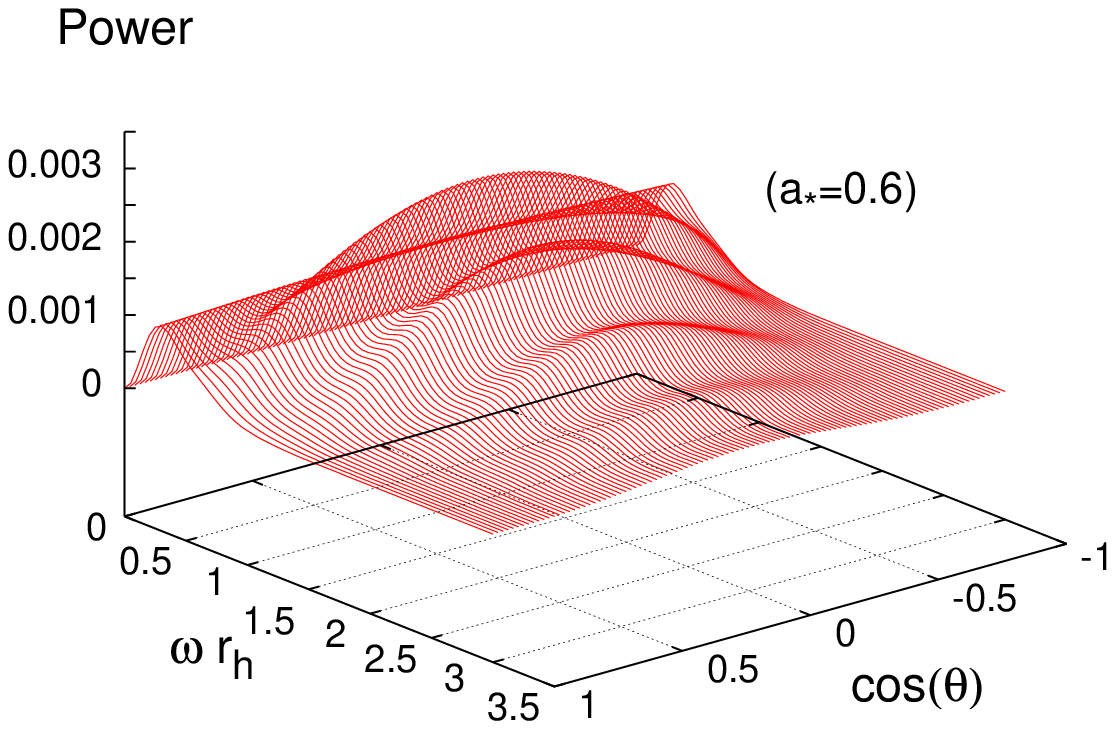, height=4.15cm}\hspace*{-1.15cm}{
\epsfig{file=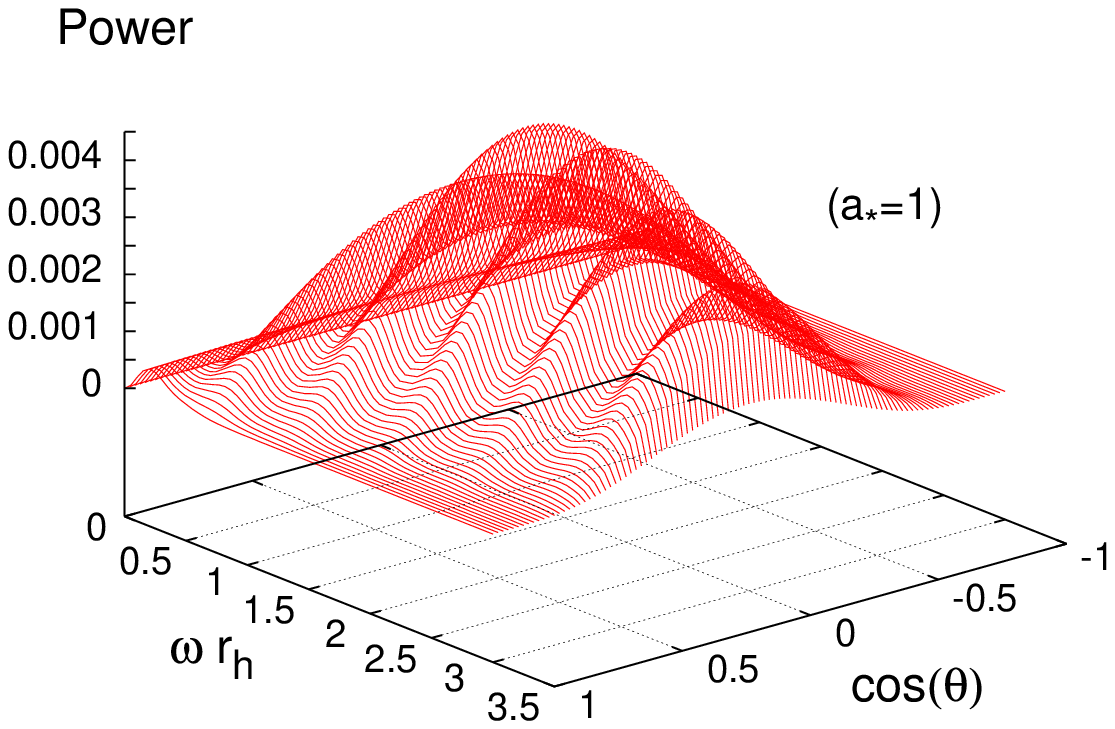, height=4.15cm}}\end{tabular}
\caption{Angular distribution of the power spectra for scalar emission from rotating
black holes, for $n=1$ and $a_*=(0,0.6,1)$.}
\label{s0n1p-ang}
\end{center}
\end{figure}

\begin{figure}[t]
\begin{center}
\begin{tabular}{c} \hspace*{-0.4cm}
\epsfig{file=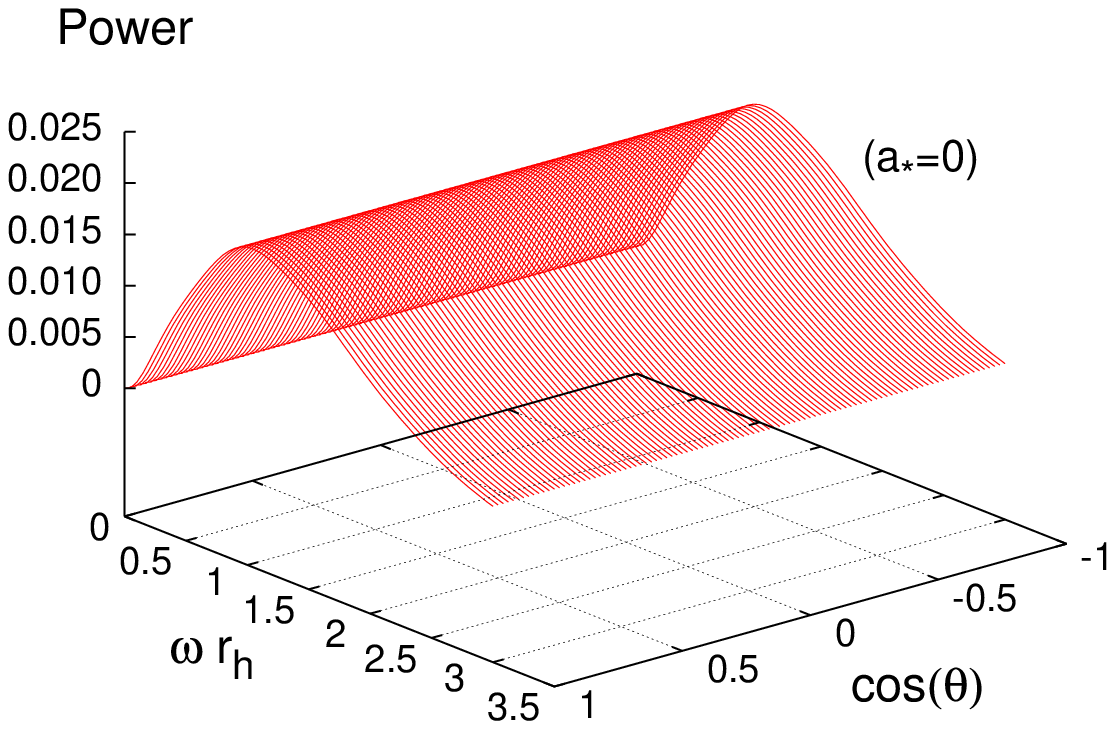, height=4.15cm}\hspace*{-1.15cm}
\epsfig{file=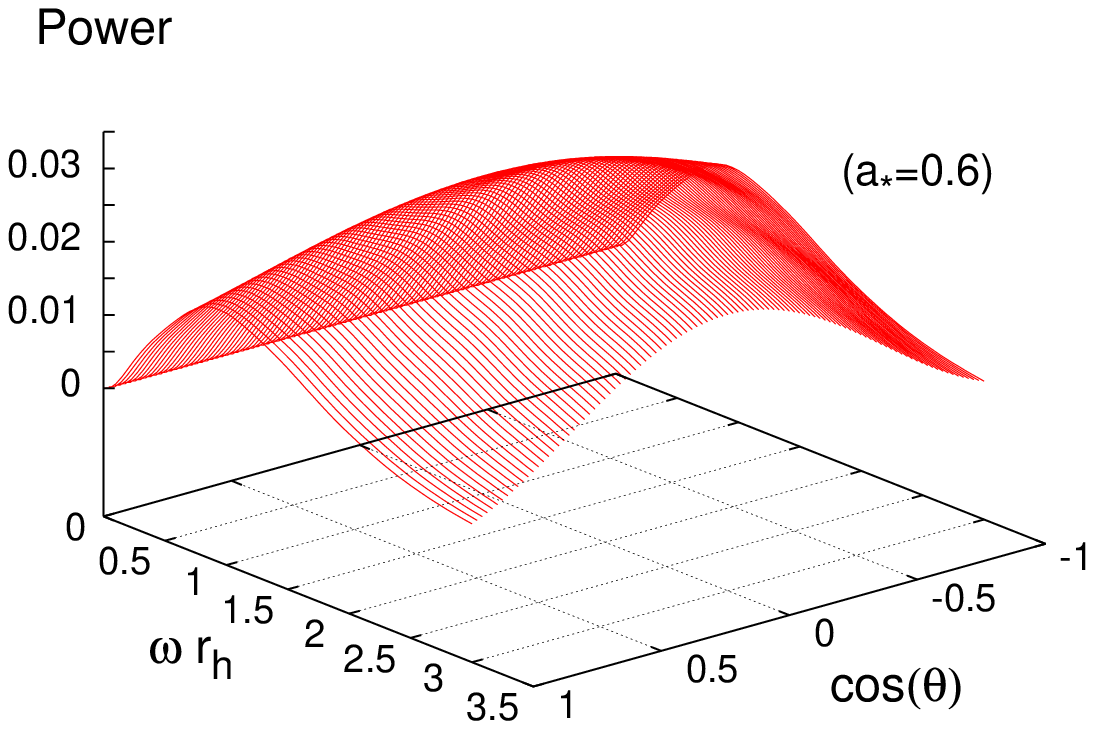, height=4.15cm}\hspace*{-1.15cm}{
\epsfig{file=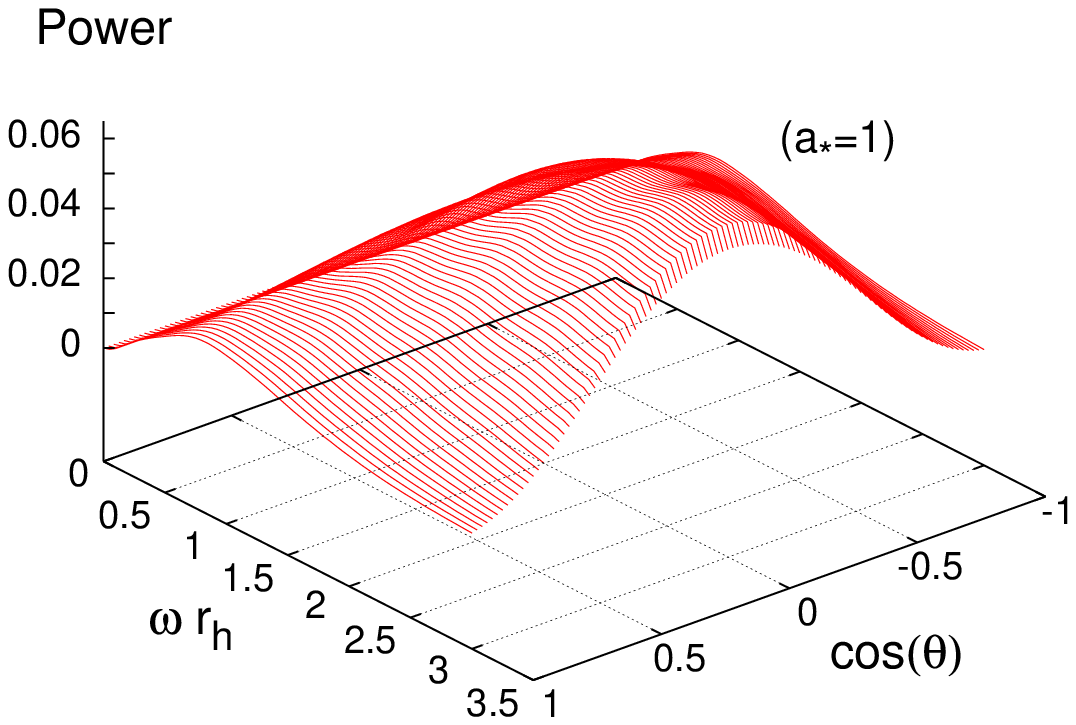, height=4.15cm}}\end{tabular}
\caption{Angular distribution of the power spectra for scalar emission from rotating
black holes, for $n=4$ and $a_*=(0,0.6,1)$.}
\label{s0n4p-ang}
\end{center}
\end{figure}

In Figs. \ref{s0n1p-ang} and \ref{s0n4p-ang}, we depict the angular distribution of the
power emission spectra from a rotating black hole on the brane for $n=1$ and $n=4$,
respectively, and for $a_*=(0,0.6,1)$. Again, for vanishing $a_*$, the spectrum shows
no angular variation, but, as the angular momentum parameter increases, the emitted modes
are starting to concentrate on the region around the equator. At low-energies, the
spectrum is again dominated by spherically symmetric behaviour, and thus remains independent
of the angle $\theta$, but gradually the non-spherically-symmetric emission becomes
important and an angular variation appears in the spectrum. As $a_*$ increases,
we can clearly observe the enhancement in the energy emission rate as well as the
shift of the peak of the curve towards higher values of $\omega r_h$. Similar behaviour,
although on a much bigger scale, can be seen as we go from the $n=1$ to the $n=4$ case.

%\smallskip
%%%%%%%%%%%%%%%
%%%%%%%%%%%%%%
%\smallskip
%%%%%%%%%%%%%%%
%%%%%%%%%%%%%%

The angular distribution of the power spectrum of a 5-dimensional
black hole ($n=1$) was also studied in Ref. \cite{IOP2}. As mentioned before, these
results were only approximate, in the sense that the
spherical harmonics were used instead of the exact spheroidal ones
(which is valid only for $a \omega \ll 1$). It would therefore be
useful to compare those results with the ones derived here, and thus check the range of
validity of the approximations made in Ref. \cite{IOP2}. By comparing our Figs.
\ref{s0n1p-ang}(a,b,c) (up to $\omega r_h=1$) with Figs. 8(a,b,c) in \cite{IOP2},
we may see that an agreement arises in general terms: as $a_*$
increases, the emitted modes start to concentrate near the equator, oscillations appear,
and the peak of the emission curve moves toward the right. Some differences though do
appear, too: (i) in our case, the spherically symmetric emission becomes sub-dominant
much faster, and the low-energy peak is rapidly suppressed, in contradiction with the
behaviour depicted in Fig. 8 of Ref. \cite{IOP2}; (ii) the height of the peaks appearing
at higher energies are also different, ours being in general lower compared
to the ones in \cite{IOP2} - this could be explained by the fact that these peaks appear
at energy values that are well beyond the range of the validity of the approximation
$a \omega \ll 1$ made in Ref. \cite{IOP2}; (iii) finally, differences appear also in the
$a_*=0$ case, with the emission curve dying out much faster in \cite{IOP2} than in our
case.

\smallskip
%%%%%%%%%%%%%%%
\begin{figure}[t]
\begin{center}
\mbox{\psfrag{x}[][][0.8]{$\omega\,r_\text{h}$}
\psfrag{y}[][][0.8]{$r_\text{h}\,d^2J/dt\,d\omega$}
\includegraphics[height=5.0cm,clip]{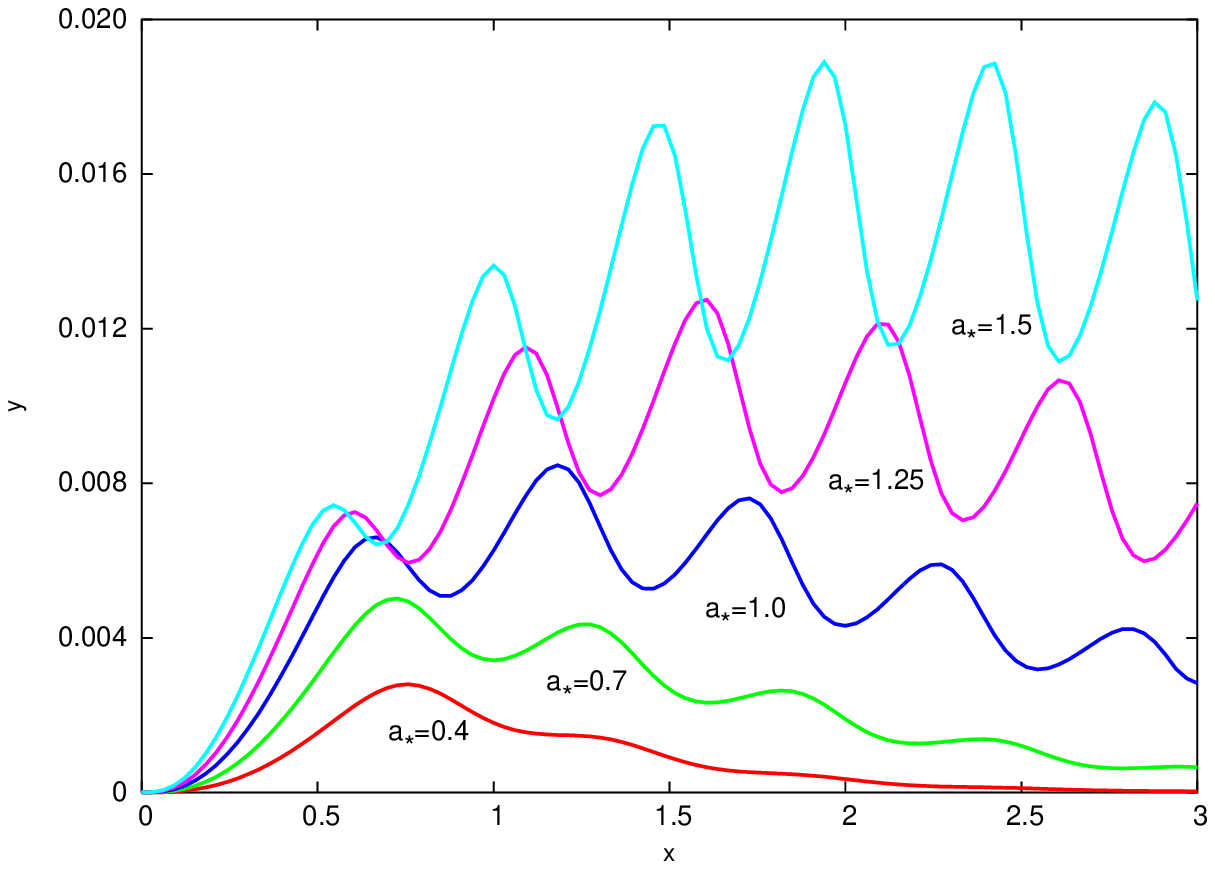}}\hspace*{0.5cm}{
\psfrag{x}[][][0.8]{$\omega\,r_\text{h}$}
\psfrag{y}[][][0.8]{$r_\text{h}\,d^2J/dt\,d\omega$}
\includegraphics[height=5.0cm,clip]{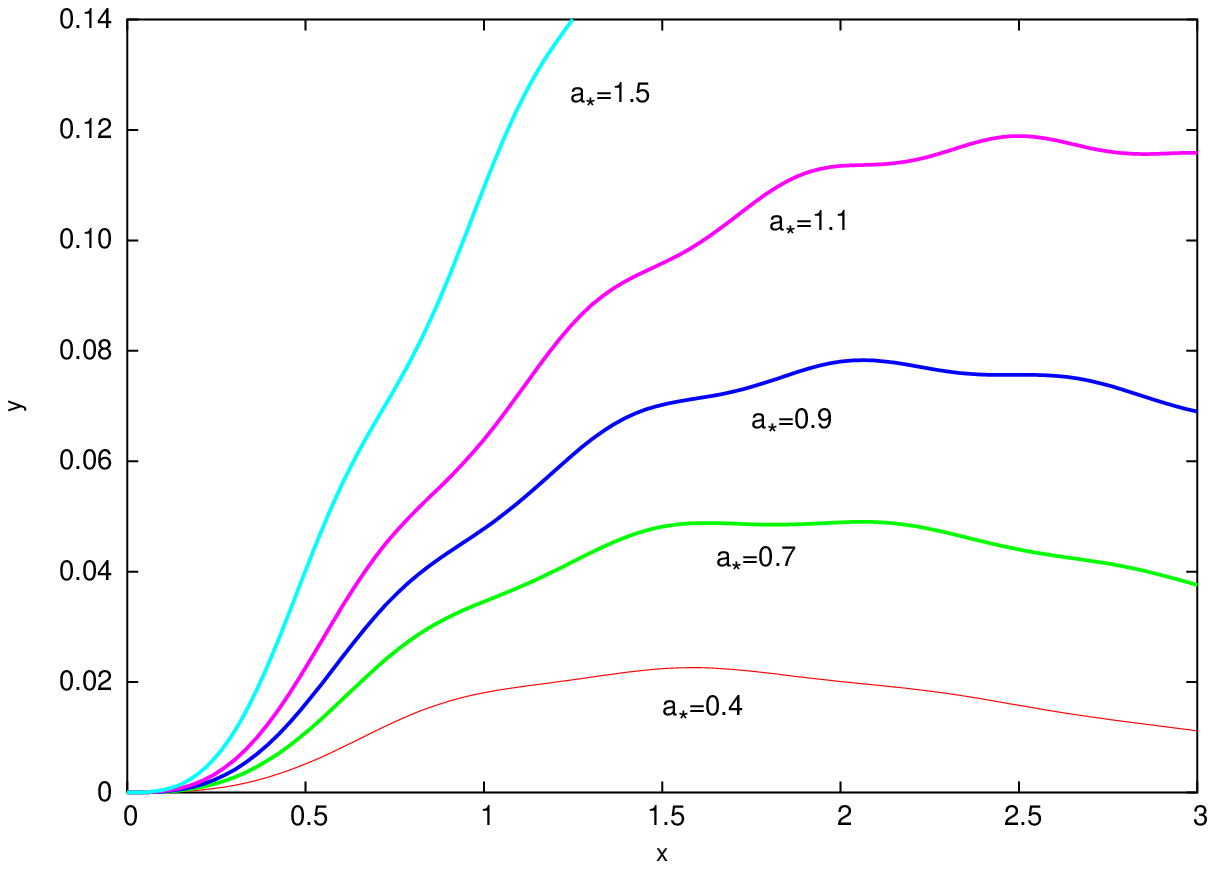}}
%\scalebox{0.8}{\rotatebox{0}{\includegraphics[width=\textwidth]{nefiopfig2.eps}}}
\caption{Angular momentum spectra for scalar particles from a rotating black hole,
for (a) $n=1$, and (b) $n=4$, and various values of $a_*$.\hspace*{1.5cm}}
\label{s0n1am}
\end{center}
\end{figure}
%%%%%%%%%%%%%%
%%%%%%%%%%%%%%%
\begin{figure}[ht]
\begin{center}
\psfrag{x}[][][0.8]{$\omega\,r_\text{h}$}
\psfrag{y}[][][0.8]{$r_\text{h}^2\,d^2J/dt\,d\omega$}
\includegraphics[height=5.5cm,clip]{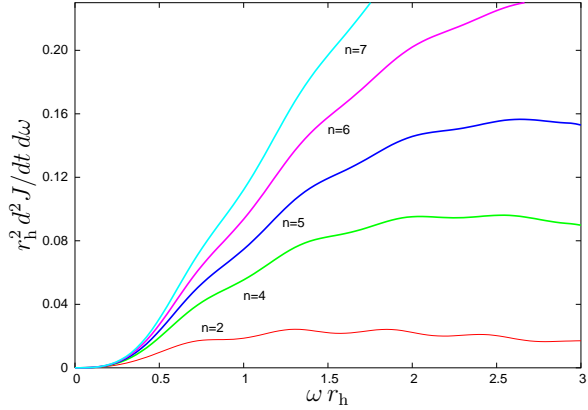}
%\scalebox{0.8}{\rotatebox{0}{\includegraphics[width=\textwidth]{nefiopfig2.eps}}}
\caption{Angular momentum spectra for scalar emission from a rotating black hole,
for $a_*=1$ and various values of $n$.\hspace*{1.5cm}}
\label{s0a1ang}
\end{center}
\end{figure}
%%%%%%%%%%%%%%%

\subsection{Angular momentum spectra}

We now turn to the rate of loss of the angular momentum of the higher-dimensional,
rotating black hole through the emission of scalar fields on the brane. In Figs.
\ref{s0n1am}(a,b), this rate is shown, again, for the indicative cases of
$n=1$ and $n=4$, respectively, and for various values of the angular momentum
parameter $a_*$. As is clear from both figures, the emission of angular
momentum is clearly enhanced at all frequencies, as $a_*$ increases, and for all
values of $n$. For
low values of $n$, oscillations are again present but, as in the case of
flux and power emission, there is an `envelope' of Gaussian shape whose
height increases and which peaks at higher values of $\omega r_{h}$, as $a_{*}$
increases. For higher values of $n$, the oscillations are smoothed out,
and the enhancement is much more significant as $a_*$ increases, due to the
substantial increase both in the height and width of the emission curve.

The dependence of the angular momentum loss rate on the number of additional
spacelike dimensions $n$ is more clearly depicted in Fig. \ref{s0a1ang}, for a
fixed value of the angular momentum parameter, i.e $a_*=1$. As in the case of
flux and power spectra, a strong enhancement can be observed in the rate of
loss of angular momentum of the black hole, as the dimensionality of spacetime
increases.

\subsection{Total emissivities}

\begin{figure}[t]
\begin{center}
\mbox{\includegraphics[height=5.5cm,clip]{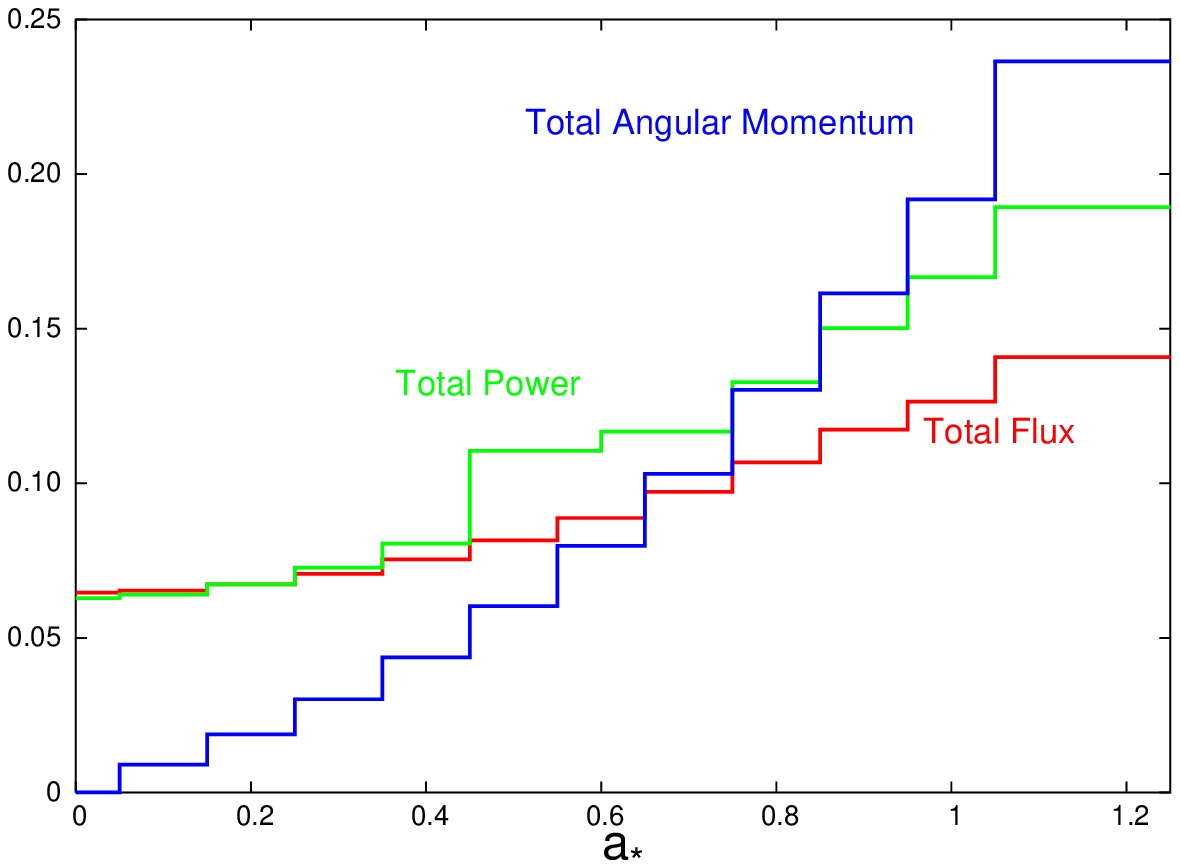}}
{\includegraphics[height=5.5cm,clip]{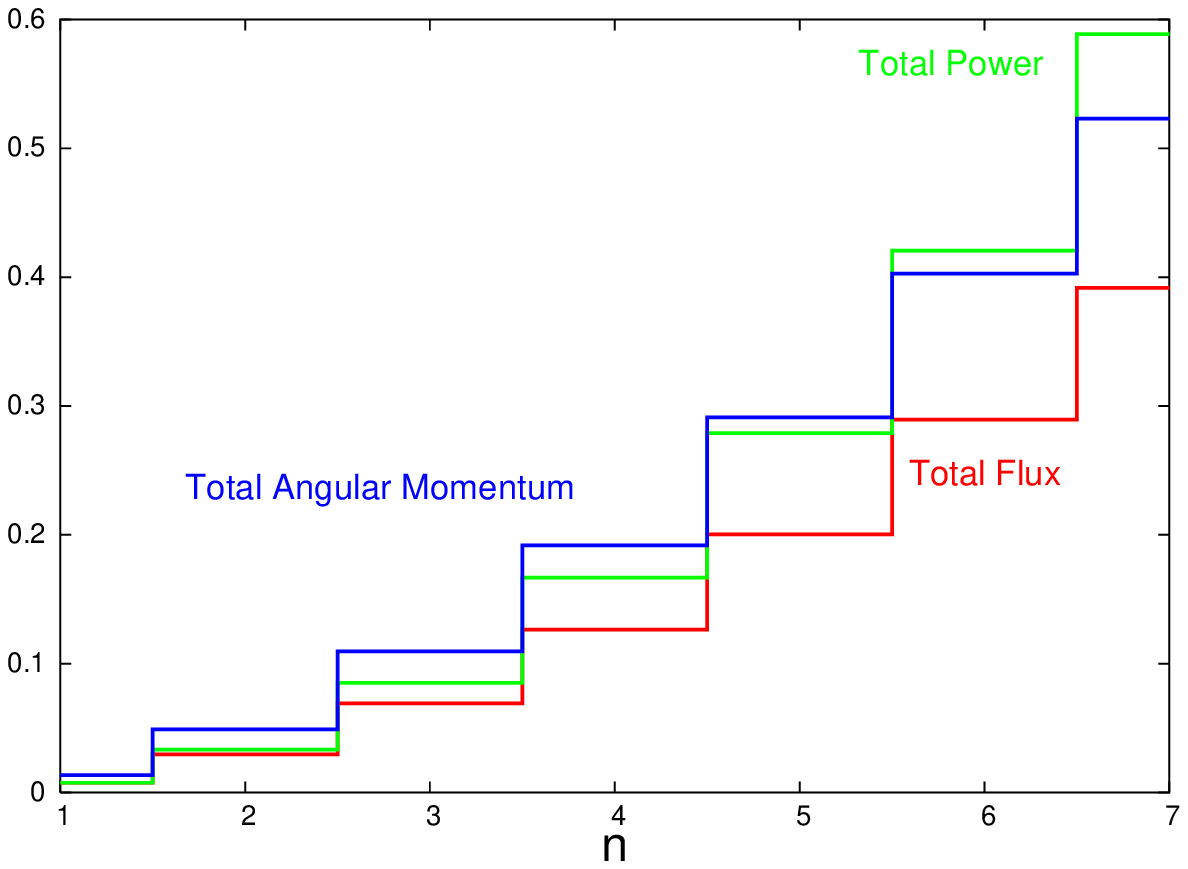}}
%\scalebox{0.8}{\rotatebox{0}{\includegraphics[width=\textwidth]{nefiopfig2.eps}}}
\caption{Total emissivities for scalar emission on the brane from a rotating black hole
as a function of (a) $a_*$, for $n=4$, and (b) $n$ for $a_*=1$.}
\label{totalna}
\end{center}
\end{figure}

We finally turn to the computation of the total emissivities of particles, energy and
angular momentum per unit time emitted by the black hole on the brane. These follow by integrating the
quantities given in Eqs. (\ref{flux})-(\ref{ang-mom}) with respect to $\omega r_{h}$,
up to the value $\omega r_{h}=3$. In Figs. \ref{totalna}(a) and (b), we present two
histograms depicting the dependence of the total fluxes on the angular momentum
parameter $a_*$ (for fixed $n=4$) and on the dimensionality of spacetime $n$ (for
fixed $a_*=1$), respectively.

As $a_*$ increases, the histogram in Fig. \ref{totalna}(a) shows that all fluxes --
particle, energy and angular momentum -- are enhanced. More specifically, for $n=4$,
as $a_*$ goes from zero to 1.25, the number of particles emitted on the brane per unit
time is enhanced by a factor of 2, the amount of energy emitted per unit time by a factor of 3,
while, as expected, the angular momentum flux (which is zero when $a_{*}=0$) increases
significantly. The same enhancement appears in all three fluxes for all values of
$n$, nevertheless the corresponding numbers are strongly $n$-dependent: for the case of
$n=1$, for instance, as $a_*$ goes again from zero to 1.25, the number of particles
emitted per unit time increases by approximately 30\%, the energy flux by a factor of 4,
while the angular momentum flux, although significantly enhanced, is an order of
magnitude smaller than the one for the $n=4$ case.

%%%%%%%%%%%%%%%
%%%%%%%%%%%%%%%

The dependence of the total emissivities on the dimensionality of spacetime is
shown in the second histogram shown in Fig. \ref{totalna}(b), that clearly illustrates
the enhancement of all fluxes as $n$ increases. For $a_*$ fixed ($a_*=1$) and $n$
varying from 1 to 7, the number of particles emitted by the black hole on the brane
per unit time is enhanced by a factor of 30, the energy flux by a factor of 100,
and the angular momentum flux by a factor of 40.

A note should be made at this point: the results presented above are only an approximation
to the exact, total emissivities that would follow by integrating the various fluxes
over the complete frequency regime. In this work, numerical results have been produced
only up to the frequency value of $\omega r_h=3$ as the integration for higher values of
$\omega r_h$ takes an unrealistically long time. For small values of $n$ and $a_*$,
the flux dies away sufficiently quickly, therefore the contribution from $\omega r_{h}>3$
is negligible and the derived emissivities are an adequate approximation to the actual
ones. However, as the value of either $a_*$ or $n$ increases, the peak of the Gaussian
curve moves to higher energies, and the contribution of the part of the spectrum that
we are missing to the exact emissivities is increasing. Nevertheless, the graphs presented
in this subsection accurately describe the correct qualitative behaviour of the exact
emissivities: the part of the spectrum that has been left out of our calculation would only increase further the enhancement of all three fluxes if it could be taken into account.

\begin{figure}[t]
\begin{center}
\mbox{\includegraphics[height=5.5cm,clip]{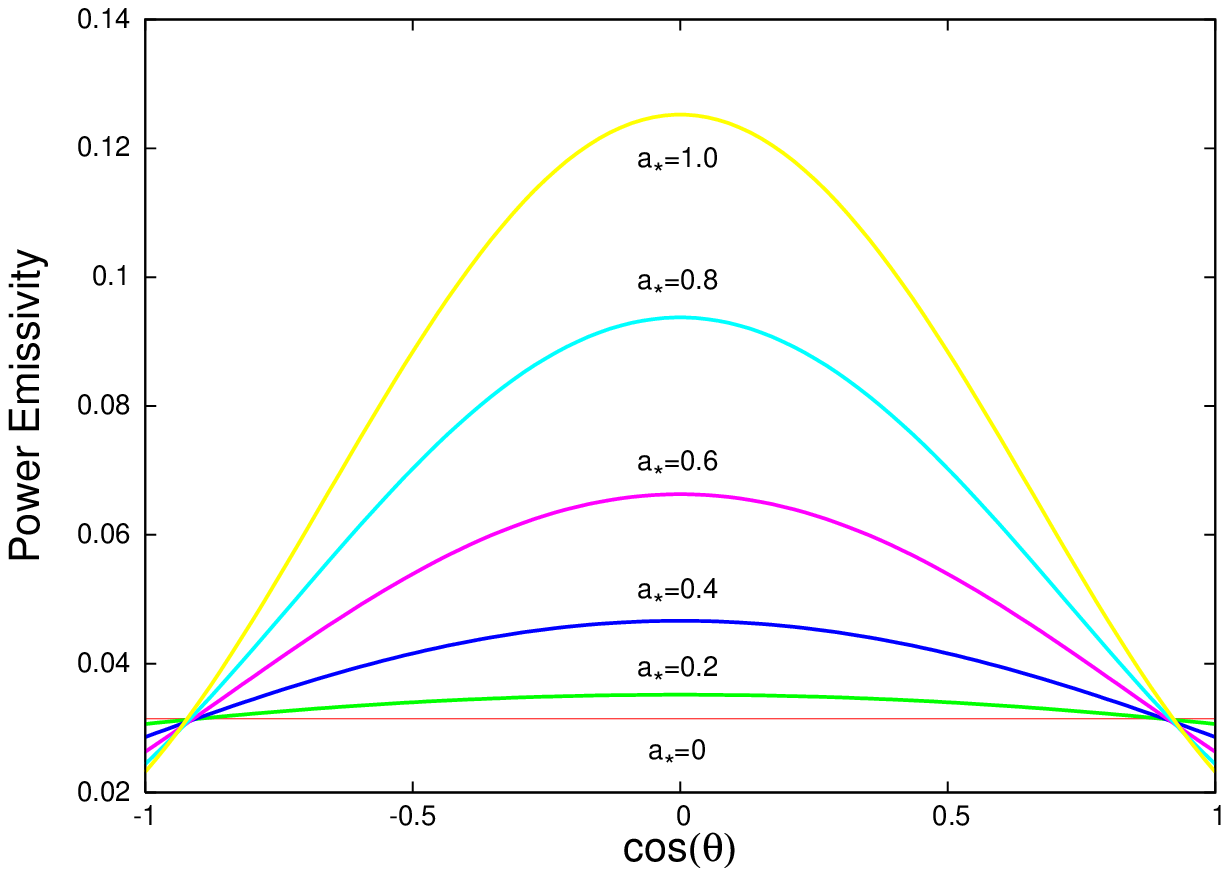}}
{\includegraphics[height=5.5cm,clip]{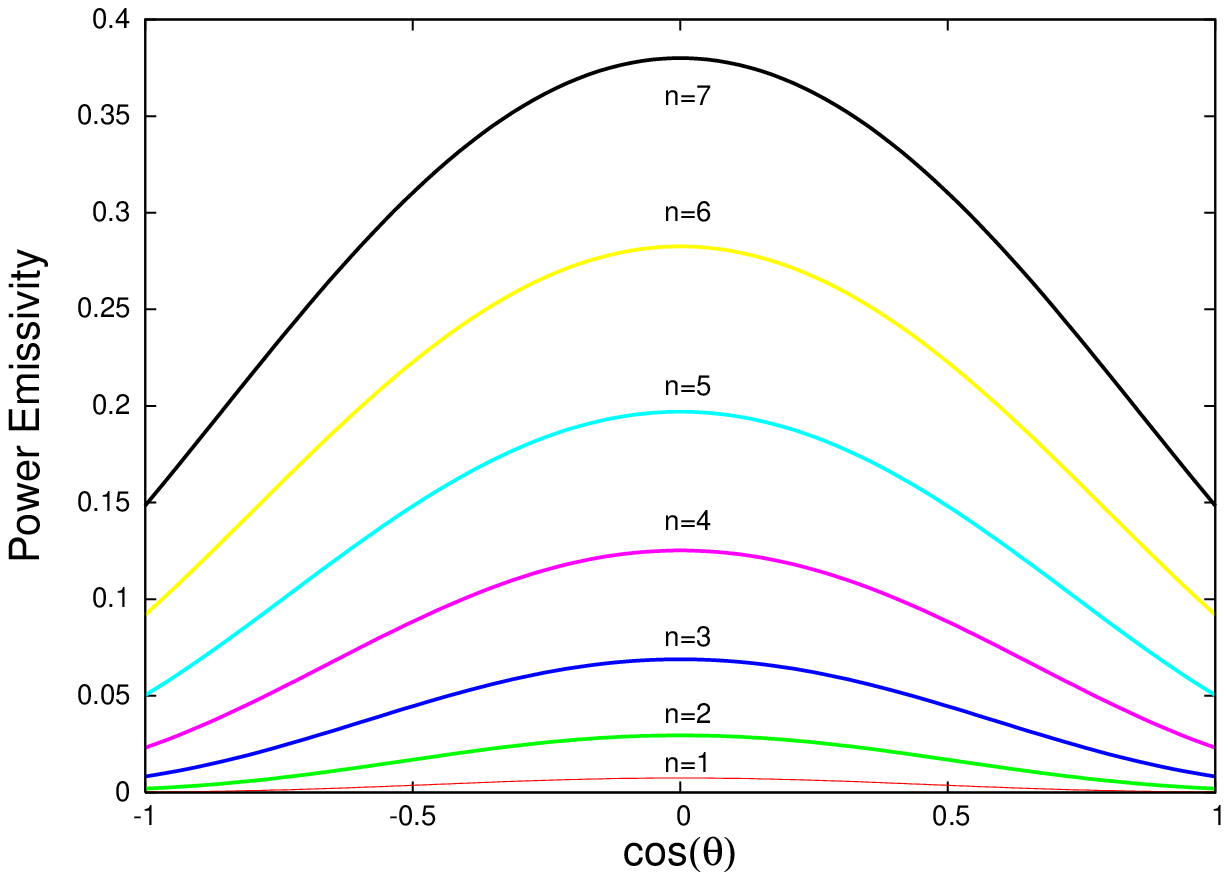}}
%\scalebox{0.8}{\rotatebox{0}{\includegraphics[width=\textwidth]{nefiopfig2.eps}}}
\caption{Power emissivity for scalar emission on the brane from a rotating black hole
as a function of $\cos \theta$, (a) for $n=4$ and variable $a_*$, and
(b) for $a_*=1$ and variable $n$.}
\label{totalna-theta}
\end{center}
\end{figure}

We have also computed the integral over the frequency of the angular distribution of the
particle flux (\ref{flux-ang}) and power spectrum (\ref{pow-ang}) to derive the
emissivity solely as a function of the azimuthal angle $\theta $. In Figs.
\ref{totalna-theta}(a,b),
we depict the power emissivity as a function of $\cos \theta$ for fixed $n$ and variable
$a_*$, and vice versa -- the flux emissivity exhibits a similar behaviour and therefore
is not shown here. The behaviour found agrees with the one deduced from the 3D graphs
of subsections 4.1 and 4.2. For fixed $n$, as the angular momentum parameter $a_*$ increases, the
energy (and particle) flux is altogether enhanced and becomes more concentrated around the
equatorial region; for fixed $a_*$, the fluxes strongly increase as $n$ increases,
and so does again the proportion of the emission concentrated in the equatorial region.

%%%%%%%%%%%%%%%
%%%%%%%%%%%%%%%

\section{Discussion and Conclusions}

In this work we have performed a comprehensive analysis of the emission of Hawking
radiation in the form of scalar fields from a $(4+n)$-dimensional, rotating black
hole on the brane. This analysis has provided complete results for the flux,
energy and angular momentum spectra for the spin-down phase of the life of a
higher-dimensional black hole, thus, filling a long-standing gap in the literature.
By means of numerical analysis, we have been able to solve both the radial and angular
equations for the scalar field modes, and our method is applicable for arbitrary
energy of the emitted particles, angular momentum of the black hole, and number
of extra, spacelike dimensions.

We have first studied the fluxes of scalar particles, energy and angular momentum --
integrated over the azimuthal angle $\theta$ -- for a range of values of the angular
momentum of the black hole $a_*$ and number of extra dimensions $n$, in each case as
a function of the mode frequency $\omega $. Due to their relation, a number of common
features arise between the flux and power spectra: for low values of $n$, both emission
rates are suppressed at the low energy regime but strongly enhanced in the intermediate-
and high-energy one, as the angular momentum parameter increases; for high values of
$n$, the enhancement is present at all frequency regimes. On the other hand, the rate
of loss of the angular momentum of the black hole increases uniformly (i.e. at all
frequency regimes) as $a_*$ increases, for all values of $n$. When the angular
momentum parameter is kept fixed and $n$ increases, a clear, strong enhancement can
be seen in all three spectra and in all frequency regimes. This enhancement can be
easily explained by the fact that, for fixed $a_{*}$, as $n$ increases, the temperature
of the black hole, given by Eq. (\ref{temperature}), increases too, thus leading to
greater emission rates.  However, for fixed $n$, increasing $a_{*}$ lowers the temperature,
and, in this case, the increased emission rates is due to enhanced super-radiance, which
speeds up the spin-down process.

We have also considered the angular distribution of the particle and energy fluxes.
Again, a number of common features arise in the two spectra. An important feature
is the dominance of the spherically-symmetric emission at the low-energy regime
that leads to an almost $\theta$-independent distribution even for a non-vanishing
angular momentum parameter; as the energy of the emitted particles increases, however,
a clear angular
dependence -- that is, the concentration of the emitted modes near the equatorial plane --
appears in both spectra. We expect the angular variation in the emission rates to be a
unique feature of the spectra during the spin-down phase.

All spectra derived above, either integrated or not over the $\theta$-coordinate,
depend strongly on the energy regime that we look at, the value of the angular momentum
parameter and the dimensionality of spacetime. In many cases, the various emission rates
change considerably as the above parameters take different, low or high, values. In
this work, care has been taken so that all relevant quantities, i.e. the angular
eigenvalue as well as the radial and angular part of the scalar field, are computed
via numerical analysis that is free from any assumptions and approximations that limit
the values of the aforementioned three, fundamental parameters of the problem.

Any assumptions made in this analysis involved only the mass of the produced black
hole, being much larger than the fundamental Planck mass, and the size of the horizon
of the black hole, being much smaller than the size of the extra dimensions. However,
the values of $M_*$ and $L$ themselves were never specified; they are free input
parameters of the problem, and thus our analysis and corresponding results are valid
for any gravitational theory with arbitrary number of extra dimensions and fundamental
scale. Finally, although our analysis assumed that the extra spacetime is empty and thus
flat, our results are also valid in the case of a Randall-Sundrum \cite{RS} type of
black hole in the limit of a horizon radius being much smaller than the AdS radius.

A number of interesting questions still remain open. Firstly, the calculation of the
rate of spin-down of the black hole \cite{Chambers:1997ai}, and thus of the duration
of this phase compared to the Schwarzschild one, requires, in addition to the
brane emission, the scalar field radiation into the bulk for all values of the number
of extra dimensions. The same calculation for the flux and energy spectra will reveal
the amount of energy lost in the bulk and thus the remaining amount available for
emission on the brane. Also, the remaining brane emission `channels', i.e. the
ones for fermions and gauge bosons, need to be investigated too. It is thus necessary,
and of great interest indeed, to extend our analysis for the emission of Hawking
radiation from a rotating, higher-dimensional black hole for higher-spin particles.
We plan to return to these questions in the near future.

{\bf Acknowledgments}.
We would like to thank Marc Casals for helpful discussions.
E.W. would like to thank the following for hospitality while this work was completed:
the Institute for Particle Physics Phenomenology, University of Durham;
the Department of Mathematical Physics, University College Dublin; and the
School of Mathematics and Statistics, University of Newcastle-upon-Tyne.
The work of P.K. was funded by the UK PPARC Research Grant PPA/A/S/2002/00350.
The work of E.W. is supported by UK PPARC, grant reference number PPA/G/S/2003/00082,
the Royal Society and the London Mathematical Society.


\begin{thebibliography}{99}

%\cite{Arkani-Hamed:1998rs}
\bibitem{ADD}
N.~Arkani-Hamed, S.~Dimopoulos and G.~R.~Dvali,
%``The hierarchy problem and new dimensions at a millimeter,''
{\it Phys.\ Lett.}\ B {\bf 429}, 263 (1998) [hep-ph/9803315];
%%CITATION = HEP-PH 9803315;%%
%``Phenomenology, astrophysics and cosmology of theories with
%sub-millimeter dimensions and TeV scale quantum gravity,''
{\it Phys.\ Rev.}\ D {\bf 59}, 086004 (1999) [hep-ph/9807344];\\
%%CITATION = HEP-PH 9807344;%%
I.~Antoniadis, N.~Arkani-Hamed, S.~Dimopoulos and G.~R.~Dvali,
%``New dimensions at a millimeter to a Fermi and superstrings at a TeV,''
{\it Phys.\ Lett.}\ B {\bf 436}, 257 (1998) [hep-ph/9804398].
%%CITATION = HEP-PH 9804398;%%

\bibitem{early}
K.~Akama,
%``An Early Proposal Of 'Brane World',''
{\it Lect.\ Notes Phys.}\  {\bf 176}, 267 (1982) [hep-th/0001113];\\
%%CITATION = HEP-TH 0001113;%%
V.~A.~Rubakov and M.~E.~Shaposhnikov,
%``Extra Space-Time Dimensions: Towards A Solution To The
% Cosmological Constant Problem,''
{\it Phys.\ Lett.}\ B {\bf 125}, 139 (1983); %%CITATION = PHLTA,B125,139;%%
%``Do We Live Inside A Domain Wall?,''
{\it Phys.\ Lett.}\ B {\bf 125}, 136 (1983);\\ %%CITATION = PHLTA,B125,136;%%
M.~Visser,
%``An Exotic Class Of Kaluza-Klein Models,''
{\it Phys.\ Lett.}\ B {\bf 159}, 22 (1985) [hep-th/9910093];\\
%%CITATION = HEP-TH 9910093;%%
G.~W.~Gibbons and D.~L.~Wiltshire,
%``Space-Time As A Membrane In Higher Dimensions,''
{\it Nucl. Phys.} {\bf B287}, 717 (1987); \\ %%CITATION = HEP-TH 0109093;%%
I.~Antoniadis,
%``A Possible New Dimension At A Few Tev,''
{\it Phys.\ Lett.}\ B {\bf 246}, 377 (1990);\\ %%CITATION = PHLTA,B246,377;%%
I.~Antoniadis, K.~Benakli and M.~Quiros,
%``Production of Kaluza-Klein states at ffuture colliders,''
{\it Phys.\ Lett.}\ B {\bf 331}, 313 (1994) [hep-ph/9403290];\\
%%CITATION = HEP-PH 9403290;%%
J.~D.~Lykken, %``Weak Scale Superstrings,''
{\it Phys.\ Rev.}\ D {\bf 54}, 3693 (1996) [hep-th/9603133].
%%CITATION = HEP-TH 9603133;%%

\bibitem{creation}
T.~Banks and W.~Fischler,
%``A model for high-energy scattering in quantum gravity,''
hep-th/9906038;\\ %%CITATION = HEP-TH 9906038;%%
D.~M.~Eardley and S.~B.~Giddings,
%{\it Classical Black Hole Production in High-Energy Collisions},
{\it Phys. Rev.} D {\bf 66}, 044011 (2002) [gr-qc/0201034];\\
%%CITATION = GR-QC 0201034;%%
H.~Yoshino and Y.~Nambu,
%{\it High-energy head-on collisions of particles and hoop conjecture},
{\it Phys. Rev.} D {\bf 66}, 065004 (2002) [gr-qc/0204060];
%%CITATION = GR-QC 0204060;%%
%``Black hole formation in the grazing collision of high-energy particles,''
{\it Phys.\ Rev.} D {\bf 67}, 024009 (2003) [gr-qc/0209003];\\
%%CITATION = GR-QC 0209003;%%
E.~Kohlprath and G.~Veneziano,
%``Black holes from high-energy beam-beam collisions,''
{\it JHEP} {\bf 0206}, 057 (2002) [gr-qc/0203093];\\ %%CITATION = GR-QC 0203093;%%
V.~Cardoso, O.~J.~C.~Dias and J.~P.~S.~Lemos,
%``Gravitational radiation in D-dimensional spacetimes,''
{\it Phys.\ Rev.}\ D {\bf 67}, 064026 (2003) [hep-th/0212168];\\
%%CITATION = HEP-TH 0212168;%%
E.~Berti, M.~Cavaglia and L.~Gualtieri,
%``Gravitational energy loss in high energy particle collisions:
%Ultrarelativistic plunge into a multidimensional black hole,''
{\it Phys. Rev.} D {\bf 69}, 124011 (2004) [hep-th/ 0309203];\\
%%CITATION = HEP-TH 0309203;%%
L.~A.~Anchordoqui, J.~L.~Feng, H.~Goldberg and A.~D.~Shapere,
%``Inelastic black hole production and large extra dimensions,''
{\it Phys.\ Lett.} B {\bf 594}, 363 (2004) [hep-ph/0311365];\\
%%CITATION = HEP-PH 0311365;%%
V.~S.~Rychkov,
%``Black hole production in particle collisions and higher curvature gravity,''
{\it Phys.\ Rev.}\ D {\bf 70}, 044003 (2004) [hep-ph/0401116];\\
%%CITATION = HEP-PH 0401116;%%
S.~B.~Giddings and V.~S.~Rychkov, %``Black holes from colliding wavepackets,''
{\it Phys.\ Rev.}\ D {\bf 70}, 104026 (2004) [hep-th/0409131];\\
%%CITATION = HEP-TH 0409131;%%
O.~I.~Vasilenko, %``Trap surface formation in high-energy black holes collision,''
hep-th/0305067; %%CITATION = HEP-TH 0305067;%%
%``Horizon formation in high-energy particles collision,''
hep-th/0407092;\\ %%CITATION = HEP-TH 0407092;%%
H.~Yoshino and V.~S.~Rychkov,
%``Improved analysis of black hole formation in high-energy particle
%collisions,''
{\it Phys.\ Rev.} D {\bf 71}, 104028 (2005) [hep-th/0503171];\\
%%CITATION = HEP-TH 0503171;%%
V.~Cardoso, E.~Berti and M.~Cavaglia,
%``What we (don't) know about black hole formation in high-energy
%collisions,''
{\it Class.\ Quant.\ Grav.}  {\bf 22}, L61 (2005) [hep-ph/0505125].
%%CITATION = HEP-PH 0505125;%%


\bibitem{colliders}
P.~C.~Argyres, S.~Dimopoulos and J.~March-Russell,
%``Black holes and sub-millimeter dimensions,''
{\it Phys.\ Lett.}\ B {\bf 441}, 96 (1998) [hep-th/9808138];\\
%%CITATION = HEP-TH 9808138;%%
R.~Emparan, G.~T.~Horowitz and R.~C.~Myers,
%{\it Black Holes Radiate Mainly on the Brane},
{\it Phys. Rev. Lett.} {\bf 85}, 499 (2000) [hep-th/0003118];\\
%%CITATION = HEP-TH 0003118;%%
S.~B.~Giddings and S.~Thomas,
%``High energy colliders as black hole factories: The end of short
%  distance physics,''
{\it Phys.\ Rev.}\ D {\bf 65}, 056010 (2002) [hep-ph/0106219];\\
%%CITATION = HEP-PH 0106219;%%
S.~Dimopoulos and G.~Landsberg, %``Black holes at the LHC,''
{\it Phys.\ Rev.\ Lett.}\  {\bf 87}, 161602 (2001) [hep-ph/0106295];\\
%%CITATION = HEP-PH 0106295;%%
S.~Dimopoulos and R.~Emparan, %``String balls at the LHC and beyond,''
{\it Phys.\ Lett.}\ B {\bf 526}, 393 (2002) [hep-ph/0108060];\\
%%CITATION = HEP-PH 0108060;%%
S.~Hossenfelder, S.~Hofmann, M.~Bleicher and H.~Stocker,
%{\it Quasi-stable black holes at LHC},
{\it Phys. Rev.} D {\bf 66}, 101502 (2002) [hep-ph/0109085];\\
%%CITATION = HEP-PH 0109085;%%
K.~Cheung, %{\it Black hole production and large extra dimensions},
{\it Phys. Rev. Lett.} {\bf 88}, 221602 (2002) [hep-ph/0110163]; \\
%%CITATION = HEP-PH 0110163;%%
%{\it Black hole, string ball, and p-brane production at hadronic
%supercolliders}
R.~Casadio and B.~Harms,
%{\it Can black holes and naked singularities be detected in accelerators?},
{\it Int. J. Mod. Phys.} A {\bf 17}, 4635 (2002) [hep-ph/0110255];\\
%%CITATION = HEP-TH 0110255;%%
S.~C.~Park and H.~S.~Song, %{\it Production of spinning black holes at colliders},
{\it J.\ Korean Phys.\ Soc.}  {\bf 43}, 30 (2003) [hep-ph/0111069];\\
%%CITATION = HEP-PH 0111069;%%
G.~Landsberg, %{\it Discovering new physics in the decays of black holes},
{\it Phys. Rev. Lett.} {\bf 88}, 181801 (2002) [hep-ph/0112061];\\
%%CITATION = HEP-PH 0112061;%%
G.~F.~Giudice, R.~Rattazzi and J.~D.~Wells,
%``Transplanckian collisions at the LHC and beyond,''
{\it Nucl.\ Phys.}\ B {\bf 630}, 293 (2002) [hep-ph/0112161];\\
%%CITATION = HEP-PH 0112161;%%
E.~J.~Ahn, M.~Cavaglia and A.~V.~Olinto, %{\it Brane factories},
{\it Phys. Lett.} B {\bf 551}, 1 (2003) [hep-th/0201042];\\
%%CITATION = HEP-TH 0201042;%%
T.~G.~Rizzo,
%{\it Black hole production at the LHC: Effects of Voloshin suppression},
{\it JHEP} {\bf 0202}, 011 (2002) [hep-ph/0201228]; hep-ph/0412087;\\
%%CITATION = HEP-PH 0412087;%%  %%CITATION = HEP-PH 0201228;%%
A.~V.~Kotwal and C.~Hays,
%{\it Production and decay of spinning black holes at colliders and tests of
%black hole dynamics},
{\it Phys. Rev.} D {\bf 66}, 116005 (2002) [hep-ph/0206055];\\
%%CITATION = HEP-PH 0206055;%%
A.~Chamblin and G.~C.~Nayak,
%{\it Black hole production at LHC: String balls and black holes from pp
%and  lead lead collisions},
{\it Phys. Rev.} D {\bf 66}, 091901 (2002) [hep-ph/0206060];\\
%%CITATION = HEP-PH 0206060;%%
T.~Han, G.~D.~Kribs and B.~McElrath,
%{ \it Black hole evaporation with separated fermions},
{\it Phys. Rev. Lett.} {\bf 90}, 031601 (2003) [hep-ph/0207003];\\
%%CITATION = HEP-PH 0207003;%%
I.~Mocioiu, Y.~Nara and I.~Sarcevic,
%{\it Hadrons as signature of black hole production at the LHC},
{\it Phys. Lett.} B {\bf 557}, 87 (2003) [hep-ph/0310073];\\
%%CITATION = HEP-PH 0301073;%%
M.~Cavaglia, S.~Das and R.~Maartens,
%{\it Will we observe black holes at LHC?},
{\it Class. Quant. Grav.} {\bf 20}, L205 (2003) [hep-ph/0305223];\\
%%CITATION = HEP-PH 0305223;%%
C.~M.~Harris, P.~Richardson and B.~R.~Webber,
%``CHARYBDIS: A black hole event generator,''
{\it JHEP} {\bf 0308}, 033 (2003) [hep-ph/0307305];\\
%%CITATION = HEP-PH 0307305;%%
R.~A.~Konoplya,
%``Gravitational quasinormal radiation of higher-dimensional black holes,''
{\it Phys.\ Rev.}\ D {\bf 68}, 124017 (2003) [hep-th/0309030];
%%CITATION = HEP-TH 0309030;%%
%``Quasinormal modes of the charged black hole in Gauss-Bonnet gravity,''
{\it Phys.\ Rev.} D {\bf 71}, 024038 (2005) [hep-th/0410057];\\
%%CITATION = HEP-TH 0410057;%%
V.~Cardoso, J.~P.~S.~Lemos and S.~Yoshida,
%``Quasinormal modes of Schwarzschild black holes in four and higher
%dimensions,''
{\it Phys.\ Rev.} D {\bf 69}, 044004 (2004) [gr-qc/0309112];\\
%%CITATION = GR-QC 0309112;%%
M.~Cavaglia and S.~Das,
%``How classical are TeV-scale black holes?,''
{\it Class.\ Quant.\ Grav.}  {\bf 21}, 4511 (2004) [hep-th/0404050];\\
%%CITATION = HEP-TH 0404050;%%
D.~Stojkovic,
%``Distinguishing between the small ADD and RS black holes in accelerators,''
{\it Phys. Rev. Lett.} {\bf 94}, 011603 (2005) [hep-ph/0409124];\\
%%CITATION = HEP-PH 0409124;%%
S.~Hossenfelder,
%``The minimal length and large extra dimensions,''
Mod.\ Phys.\ Lett.\ A {\bf 19}, 2727 (2004) [hep-ph/0410122];\\
%%CITATION = HEP-PH 0410122;%%
C.~M.~Harris, M.~J.~Palmer, M.~A.~Parker, P.~Richardson, A.~Sabetfakhri
and B.~R.~Webber,
%``Exploring higher dimensional black holes at the Large Hadron Collider,''
hep-ph/0411022;\\
%%CITATION = HEP-PH 0411022;%%
T.~G.~Rizzo,
%``Warped phenomenology of higher-derivative gravity,''
{\it JHEP} {\bf 0501}, 028 (2005) [hep-ph/0412087];
%%CITATION = HEP-PH 0412087;%%
%``Collider production of TeV scale black holes and higher-curvature
%gravity,''
{\it JHEP} {\bf 0506}, 079 (2005) [hep-ph/0503163];\\
%%CITATION = HEP-PH 0503163;%%
J.~L.~Hewett, B.~Lillie and T.~G.~Rizzo, hep-ph/0503178;\\
%``Black holes in many dimensions at the LHC: Testing critical string
%theory,''
%%CITATION = HEP-PH 0503178;%%
L.~Lonnblad, M.~Sjodahl and T.~Akesson, hep-ph/0505181;\\
%``QCD-supression by black hole production at the LHC,''
%%CITATION = HEP-PH 0505181;%%
B.~Koch, M.~Bleicher and S.~Hossenfelder, hep-ph/0507138; hep-ph/0507140.
%``Black hole remnants at the LHC,''
%%CITATION = HEP-PH 0507138;%%
%``Trapping black hole remnants,''
%%CITATION = HEP-PH 0507140;%%

\bibitem{cosmic} A.~Goyal, A.~Gupta and N.~Mahajan,
%{\it Neutrinos as source of ultra high-energy cosmic rays in extra
%dimensions},
{\it Phys. Rev.} D {\bf 63}, 043003 (2001) [hep-ph/0005030];\\
%%CITATION = HEP-PH 0005030;%%
J.~L.~Feng and A.~D.~Shapere, %{\it Black hole production by cosmic rays},
{\it Phys. Rev. Lett.} {\bf 88}, 021303 (2002) [hep-ph/0109106];\\
%%CITATION = HEP-PH 0109106;%%
L.~Anchordoqui and H.~Goldberg,
%{\it Experimental signature for black hole production in neutrino air
%showers},
{\it Phys. Rev.} D {\bf 65}, 047502 (2002) [hep-ph/0109242];\\ %%CITATION = HEP-PH 0109242;%%
R.~Emparan, M.~Masip and R.~Rattazzi,
%{\it Cosmic rays as probes of large extra dimensions and TeV gravity},
{\it Phys. Rev.} D {\bf 65}, 064023 (2002) [hep-ph/ 0109287];\\ %%CITATION = HEP-PH 0109287;%%
L.~A.~Anchordoqui, J.~L.~Feng, H.~Goldberg and A.~D.~Shapere,
%``Black holes from cosmic rays: Probes of extra dimensions and new limits
%  on TeV-scale gravity,''
{\it Phys.\ Rev.}\ D {\bf 65}, 124027 (2002) [hep-ph/0112247]; {\it Phys. Rev.} D {\bf 68},
104025 (2003) [hep-ph/0307228];\\ %%CITATION = HEP-PH 0307228;%%
%%CITATION = HEP-PH 0112247;%%
Y.~Uehara,
%``Production and detection of black holes at neutrino array,''
{\it Prog.\ Theor.\ Phys.}\  {\bf 107}, 621 (2002) [hep-ph/0110382];\\
%%CITATION = HEP-PH 0110382;%%
J.~Alvarez-Muniz, J.~L.~Feng, F.~Halzen, T.~Han and D.~Hooper,
%``Detecting microscopic black holes with neutrino telescopes,''
{\it Phys.\ Rev.}\ D {\bf 65}, 124015 (2002) [hep-ph/0202081];\\
%%CITATION = HEP-PH 0202081;%%
A.~Ringwald and H.~Tu,
%``Collider versus cosmic ray sensitivity to black hole production,''
{\it Phys.\ Lett.}\ B {\bf 525}, 135 (2002) [hep-ph/0111042];\\
%%CITATION = HEP-PH 0111042;%%
M.~Kowalski, A.~Ringwald and H.~Tu,
%``Black holes at neutrino telescopes,''
{\it Phys.\ Lett.}\ B {\bf 529}, 1 (2002) [hep-ph/0111042];\\ %%CITATION = HEP-PH 0201139;%%
E.~J.~Ahn, M.~Ave, M.~Cavaglia and A.~V.~Olinto,
%``TeV black hole fragmentation and detectability in extensive  air-showers,''
{\it Phys. Rev.} D {\bf 68}, 043004 (2003) [hep-ph/0306008];\\
%%CITATION = HEP-PH 0306008;%%
E.~J.~Ahn, M.~Cavaglia and A.~V.~Olinto,
%``Uncertainties in limits on TeV-gravity from neutrino induced air showers,''
{\it Astropart.\ Phys.}  {\bf 22}, 377 (2005) [hep-ph/0312249];\\
%%CITATION = HEP-PH 0312249;%%
T.~Han and D.~Hooper,
%``The particle physics reach of high-energy neutrino astronomy,''
{\it New J.\ Phys.}  {\bf 6}, 150 (2004) [hep-ph/0408348];\\
%%CITATION = HEP-PH 0408348;%%
A.~Cafarella, C.~Coriano and T.~N.~Tomaras,
%``Cosmic ray signals from mini black holes in models with extra dimensions: An
%analytical / Monte Carlo study,''
hep-ph/0410358;\\
%%CITATION = HEP-PH 0410358;%%
A.~Barrau, C.~Feron and J.~Grain, astro-ph/0505436.
%``Astrophysical production of microscopic black holes in a low Planck-scale
%world,''
%%CITATION = ASTRO-PH 0505436;%%



\bibitem{Kanti} P.~Kanti,
%``Black holes in theories with large extra dimensions: A review,''
{\it Int. J. Mod. Phys.} A {\bf 19}, 4899 (2004) [hep-ph/0402168].
%%CITATION = HEP-PH 0402168;%%

\bibitem{reviews} M.~Cavaglia,
%``Black hole and brane production in TeV gravity: A review,''
{\it Int. J. Mod. Phys.} A {\bf 18}, 1843 (2003) [hep-ph/0210296];\\
%%CITATION = HEP-PH 0210296;%%
G.~Landsberg, %``Black holes at future colliders and in cosmic rays,''
{\it Eur. Phys. J.} C {\bf 33}, S927 (2004) [hep-ex/0310034];\\
%%CITATION = HEP-EX 0310034;%%
K.~Cheung, hep-ph/0409028;\\
%``Collider phenomenology for a few models of extra dimensions,''
%%CITATION = HEP-PH 0409028;%%
S.~Hossenfelder, hep-ph/0412265;\\
%``What Black Holes Can Teach Us,''
%%CITATION = HEP-PH 0412265;%%
A.~S.~Majumdar and N.~Mukherjee, astro-ph/0503473.
%``Braneworld black holes in cosmology and astrophysics,''
%%CITATION = ASTRO-PH 0503473;%%

\bibitem{Harris} C.~M. Harris, hep-ph/0502005.
%``Physics beyond the standard model: Exotic leptons and black holes at future
%colliders,''
%%CITATION = HEP-PH 0502005;%%


\bibitem{kmr1}
P.~Kanti and J.~March-Russell,
%``Calculable corrections to brane black hole decay. I: The scalar case,''
{\it Phys.\ Rev.}\ D {\bf 66}, 024023 (2002) [hep-ph/0203223].
%%CITATION = HEP-PH 0203223;%%

\bibitem{Frolov1}
V.~P.~Frolov and D.~Stojkovic,
%``Black hole radiation in the brane world and recoil effect,''
{\it Phys.\ Rev.}\ D {\bf 66}, 084002 (2002) [hep-th/0206046].
%%CITATION = HEP-TH 0206046;%%

\bibitem{kmr2} P.~Kanti and J.~March-Russell,
%``Calculable corrections to brane black hole decay. II: Greybody factors for
%spin 1/2 and 1,''
{\it Phys.\ Rev.}\ D {\bf 67}, 104019 (2003) [hep-ph/0212199].
%%CITATION = HEP-PH 0212199;%%

\bibitem{HK1} C.~M.~Harris and P.~Kanti,
%``Hawking radiation from a (4+n)-dimensional black hole: Exact results  for
%the Schwarzschild phase,''
{\it JHEP} {\bf 0310}, 014 (2003) [hep-ph/0309054].
%%CITATION = HEP-PH 0309054;%%

\bibitem{Barrau} A.~Barrau, J.~Grain and S.~O.~Alexeyev,
%``Gauss-Bonnet black holes at the LHC: Beyond the dimensionality of space,''
{\it Phys. Lett.} B {\bf 584}, 114 (2004). %%CITATION = HEP-PH 0311238;%%

\bibitem{Jung-mass}
E.~l.~Jung, S.~H.~Kim and D.~K.~Park,
%``Absorption cross section for S-wave massive scalar,''
{\it Phys.\ Lett.} B {\bf 586}, 390 (2004) [hep-th/0311036];
%%CITATION = HEP-TH 0311036;%%
%``Low-energy absorption cross section for massive scalar and Dirac fermion by
%(4+n)-dimensional Schwarzschild black hole,''
{\it JHEP} {\bf 0409}, 005 (2004) [hep-th/0406117];
%%CITATION = HEP-TH 0406117;%%
%``Proof of universality for the absorption of massive scalar by the
%higher-dimensional Reissner-Nordstroem black holes,''
{\it Phys.\ Lett.}\ B {\bf 602}, 105 (2004) [hep-th/0409145].
%%CITATION = HEP-TH 0409145;%%

\bibitem{Doran}
M.~Doran and J.~Jaeckel, astro-ph/0501437.
%``Testing dark energy and light particles via black hole evaporation at
%colliders,''
%%CITATION = ASTRO-PH 0501437;%%

\bibitem{Jung-charge}
E.~Jung and D.~K.~Park,
%``Absorption and emission spectra of an higher-dimensional
%Reissner-Nordstroem black hole,''
{\it Nucl.\ Phys.}\ B {\bf 717}, 272 (2005) [hep-th/0502002];\\
%%CITATION = HEP-TH 0502002;%%
E.~Jung, S.~Kim and D.~K.~Park,
%``Ratio of absorption cross section for Dirac fermion to that for scalar in
%the higher-dimensional black hole background,''
{\it Phys.\ Lett.} B {\bf 614}, 78 (2005)  [hep-th/0503027].
%%CITATION = HEP-TH 0503027;%%

\bibitem{BGK}
P.~Kanti, J.~Grain and A.~Barrau,
%``Bulk and brane decay of a (4+n)-dimensional Schwarzschild-De-Sitter black
%hole: Scalar radiation,''
{\it Phys.\ Rev.} D {\bf 71}, 104002 (2005) [hep-th/0501148].
%%CITATION = HEP-TH 0501148;%%

\bibitem{Frolov2}
V.~P.~Frolov and D.~Stojkovic,
%``Quantum radiation from a 5-dimensional rotating black hole,''
{\it Phys.\ Rev.}\ D {\bf 67}, 084004 (2003) [gr-qc/0211055].
%%CITATION = GR-QC 0211055;%%

\bibitem{IOP1}
D.~Ida, K.~y.~Oda and S.~C.~Park,
%``Rotating black holes at future colliders: Greybody factors for
%  brane fields,''
{\it Phys.\ Rev.}\ D {\bf 67}, 064025 (2003), {\it Erratum-ibid.}\ D {\bf 69},
049901 (2004) [hep-th/0212108].
%%CITATION = HEP-TH 0212108;%%

\bibitem{HK2}
C.~M.~Harris and P.~Kanti,
%``Hawking radiation from a (4+n)-dimensional rotating black hole,''
hep-th/0503010.
%%CITATION = HEP-TH 0503010;%%

\bibitem{IOP-proc}
D.~Ida, K.~y.~Oda and S.~C.~Park, hep-ph/0501210.
%``Anisotropic scalar field emission from TeV scale black hole,''
%%CITATION = HEP-PH 0501210;%%

\bibitem{Jung-super} E.~Jung, S.~Kim and D.~K.~Park,
%``Condition for superradiance in higher-dimensional rotating black holes,''
{\it Phys.\ Lett.} B {\bf 615}, 273 (2005) [hep-th/0503163]; hep-th/0504139.
%%CITATION = HEP-TH 0503163;%%
%``Condition for the superradiance modes in higher-dimensional rotating black
%holes with multiple angular momentum parameters,''
%%CITATION = HEP-TH 0504139;%%

\bibitem{IOP2}
D.~Ida, K.~y.~Oda and S.~C.~Park,
{\it Phys.\ Rev.} D {\bf 71}, 124039 (2005)
[hep-th/0503052].
%``Rotating black holes at future colliders. II: Anisotropic scalar field
%emission,''
%%CITATION = HEP-TH 0503052;%%

\bibitem{Nomura} H.~Nomura, S.~Yoshida, M.~Tanabe and K.~i.~Maeda,
%``The fate of a five-dimensional rotating black hole via Hawking radiation,''
hep-th/0502179.
%%CITATION = HEP-TH 0502179;%%


\bibitem{Jung-rot}E.~Jung and D.~K.~Park, hep-th/0506204.
%``Bulk versus brane in the absorption and emission: 5D rotating black hole
%case,''
%%CITATION = HEP-TH 0506204;%%

\bibitem{CKW} M. Casals, P. Kanti and E. Winstanley, in progress.

\bibitem{MP}
R.~C.~Myers and M.~J.~Perry,
%``Black Holes In Higher Dimensional Space-Times,''
{\it Annals Phys.}\  {\bf 172}, 304 (1986).
%%CITATION = APNYA,172,304;%%

\bibitem{Hawking}
S.~W.~Hawking,
%``Particle Creation By Black Holes,''
{\it Commun.\ Math.\ Phys.}\  {\bf 43}, 199 (1975).
%%CITATION = CMPHA,43,199;%%


\bibitem{classics} W.~T.~Zaumen, {\it Nature} {\bf 247}, 530 (1974);\\
B.~Carter, %``Charge And Particle Conservation In Black Hole Decay,''
{\it Phys. Rev. Lett.} {\bf 33}, 558 (1974);\\ %%CITATION = PRLTA,33,558;%%
A.~A.~Starobinskii and S.~M.~Churilov, {\it Sov. Phys.-JETP} {\bf 38}, 1 (1974);\\
G.~W.~Gibbons, {\it Commun. Math. Phys.} {\bf 44}, 245 (1975);\\
W.~G.~Unruh, %``Absorption Cross-Section Of Small Black Holes,''
{\it Phys.\ Rev.}\ D {\bf 14}, 3251 (1976).%%CITATION = PHRVA,D14,3251;%%

\bibitem{page} D.~N.~Page,
%{\it Particle Emission Rates from a Black Hole: Massless Particles from
%an Uncharged, Nonrotating Hole},
{\it Phys. Rev.} D {\bf 13}, 198 (1976); %%CITATION = PHRVA,D13,198;%%
%{\it Dirac Equation around a charged rotating black hole},
{\it Phys. Rev.} D {\bf 14}, 1509 (1976); %%CITATION = PHRVA,D14,1509;%%
%{\it Particle Emission Rates from a Black Hole. II. Massless Particles
%from a rotating Hole},
{\it Phys. Rev.} D {\bf 14}, 3260 (1976); %%CITATION = PHRVA,D14,3260;%%
%{\it Particle Emission Rates from a Black Hole. III. Charged leptons
%from a nonrotating hole},
{\it Phys. Rev.} D {\bf 16}, 2402 (1977). %%CITATION = PHRVA,D16,2402;%%

\bibitem{sanchez} N.~Sanchez,
%{\it Wave Scattering theory and the absorption problem for a black hole},
{\it Phys. Rev.} D {\bf 16}, 937 (1977); %%CITATION = PHRVA,D16,937;%%
%{\it Absorption And Emission Spectra Of A Schwarzschild Black Hole},
{\it Phys. Rev.} D {\bf 18}, 1030 (1978); %%CITATION = PHRVA,D18,1030;%%
%{\it Elastic Scattering of Waves by a Black Hole},
{\it Phys. Rev.} D {\bf 18}, 1798 (1978). %%CITATION = PHRVA,D18,1798;%%

\bibitem{spheroidals}
M.~Abramowitz and I.~A.~Stegun, {\it {Handbook of Mathematical Functions}}
(Dover, New York, 1964);\\
C.~Flammer, {\it {Spheroidal wave functions}} (Stanford University Press, 1957);\\
J.~N.~Goldberg, A.~J.~MacFarlane, E.~T.~Newman, F.~Rohrlich and E.~C.~Sudarshan,
%{\it Spin S Spherical Harmonics And Edth},
{\it J. Math. Phys.} {\bf 8}, 2155 (1967);\\ %%CITATION = JMAPA,8,2155;%%
J.~Meixner, F.~W.~Sch\"afke, and G.~Wolf,
{\it {Mathieu functions and spheroidal functions and their mathematical foundations, further studies}} (Springer-Verlag, Berlin, 1980);\\
E.~Seidel,
%``A Comment On The Eigenvalues Of Spin Weighted Spheroidal Functions,''
{\it Class.\ Quant.\ Grav.} {\bf 6}, 1057 (1989). %%CITATION = CQGRD,6,1057;%%

\bibitem{ottewill}
A.~C.~Ottewill and E.~Winstanley,
%``The renormalized stress tensor in Kerr space-time: General results,''
{\it Phys.\ Rev.} D {\bf 62}, 084018 (2000) [gr-qc/0004022].
%%CITATION = GR-QC 0004022;%%

\bibitem{fandt}
V.~P.~Frolov and K.~S.~Thorne,
%``Renormalized Stress - Energy Tensor Near The Horizon Of A Slowly Evolving,
%Rotating Black Hole,''
{\it Phys.\ Rev.} D {\bf 39}, 2125 (1989).
%%CITATION = PHRVA,D39,2125;%%

\bibitem{gavin}
G.~Duffy, PhD thesis, University College Dublin (2002).
\newline
G.~Duffy and A.~C.~Ottewill, to appear.

\bibitem{Wasserstrom}
E. Wasserstrom, {\it J. Comp. Phys.} {\bf 9}, 53 (1972).

\bibitem{NR}
W.~H.~Press, S.~A.~Teukolsky, W.~T.~Vetterling and B.~P.~Flannery,
{\it {Numerical Recipes in Fortran}}
(Cambridge University Press, 1986).

\bibitem{marc}
M.~Casals and A.~C.~Ottewill,
%``High Frequency Asymptotics for the Spin-Weighted Spheroidal Equation,''
{\it Phys.\ Rev.} D {\bf 71}, 064025 (2005) [gr-qc/0409012].
%%CITATION = GR-QC 0409012;%%

\bibitem{leaver}
E.~W.~Leaver, Proc.\ Roy.\ Soc.\ London A {\bf {402}}, 285 (1985).


\bibitem{RS} L.~Randall and R.~Sundrum,
%{\it A Large Mass Hierarchy from a Small Extra Dimension},
{\it Phys. Rev. Lett.} {\bf 83}, 3370 (1999) [hep-ph/9905221];
%%CITATION = HEP-PH 9905221;%%
%{\it An alternative to compactification},
{\it Phys. Rev. Lett.} {\bf 83}, 4690 (1999) [hep-ph/9906064]. %%CITATION = HEP-TH 9906064;%%

%\cite{Chambers:1997ai}
\bibitem{Chambers:1997ai}
C.~M.~Chambers, W.~A.~Hiscock and B.~Taylor,
%``Spinning down a black hole with scalar fields,''
{\it Phys.\ Rev.\ Lett.}  {\bf 78}, 3249 (1997) [gr-qc/9703018].
%%CITATION = GR-QC 9703018;%%

%\bibitem{gavin}
%Gavin Duffy, PhD thesis, University College Dublin, 2002.





\end{thebibliography}
\end{document}